\documentclass[aps,prd,reprint,onecolumn,nofootinbib,superscriptaddress]{revtex4-2}

\usepackage{dcolumn}
\usepackage{bm}
\usepackage[latin1]{inputenc}
\usepackage[spanish,english]{babel}
\usepackage{amsfonts}
\usepackage{amssymb}
\usepackage{graphicx}
\usepackage{setspace}
\usepackage{verbatim} 
\usepackage{xcolor}
\usepackage{hyperref}
\usepackage{mathrsfs}
\usepackage{cancel}
\usepackage{amsmath}
\usepackage{braket}
\usepackage{color}
\usepackage{soul} 
\usepackage{slashed} 
\usepackage[normalem]{ulem} 

\numberwithin{equation}{section}

\newcommand{\be}{\begin{equation}}
\newcommand{\ee}{\end{equation}}
\newcommand{\bea}{\begin{eqnarray}}
\newcommand{\eea}{\end{eqnarray}}

\colorlet{RED}{red}
\colorlet{ORANGE}{orange}

\newcommand{\scri}{\mathcal{I}}

\begin{document}

\title{Singularity resolution and unitarity in two-dimensional dilaton black holes with negative central charge}

\author{C\'esar Garc\'ia-P\'erez}
\email{cesar.garcia.perez@edu.unige.it} 
\affiliation{Instituto de F\'isica Corpuscular (IFIC), CSIC-Universitat de Val\`encia and Departament de F\'isica Te\`orica, Facultad de F\'isica,  Burjassot-46100, Valencia, Spain}
\affiliation{DIME, Universit\`a di Genova, Via all'Opera Pia 15, 16145 Genova, Italy}
\affiliation{INFN Sezione di Genova, Via Dodecaneso 33, 16146 Genova, Italy}

\author{F. Javier Mara\~{n}\'on-Gonz\'alez}
\email{jmarag@ific.uv.es}
\affiliation{Instituto de F\'isica Corpuscular (IFIC), CSIC-Universitat de Val\`encia and Departament de F\'isica Te\`orica, Facultad de F\'isica,  Burjassot-46100, Valencia, Spain}

\author{Jos\'e Navarro-Salas}
\email{jnavarro@ific.uv.es} 
\affiliation{Instituto de F\'isica Corpuscular (IFIC), CSIC-Universitat de Val\`encia and Departament de F\'isica Te\`orica, Facultad de F\'isica,  Burjassot-46100, Valencia, Spain}

\author{Silvia Pla}
\email{silvia.pla-garcia@tum.de}
\affiliation{Physik-Department, Technische Universit\"at M\"unchen, James-Franck-Str., 85748 Garching, Germany}

\begin{abstract}

We study a one-loop corrected extension of the classical Callan--Giddings--Harvey--Strominger (CGHS) model of two-dimensional dilaton gravity. The effective action combines the non-local Polyakov action for matter fluctuations, a Polyakov-type term built from an auxiliary flat metric that implements Strominger's mechanism for the Faddeev--Popov reparametrization ghosts, and a local counterterm that simultaneously preserves the flatness of the auxiliary metric, ensures exact solvability, and keeps two-dimensional Minkowski spacetime as an exact solution of the backreacted equations. In the regime of negative total central charge, the classical curvature singularity is resolved and gives way to asymptotically flat regions inside the  horizon. The exterior Hawking flux is preserved and turns out to be correlated with an internal radiation flux supported on null surfaces that approach null infinity from beyond the horizon; this internal flux, in particular, presents a short interval of negative values. These correlations point to the preservation of unitarity, provided the relevant null surfaces remain at a finite affine distance from the collapsing matter trajectory.  Within the present formulation, however,  a fully consistent energy balance cannot yet be established. We discuss possible strategies to overcome this issue.

\end{abstract}

\maketitle

\tableofcontents

\section{Introduction}

The question of unitary evolution of black holes,  first formulated by Hawking nearly fifty years ago \cite{Hawking76}, remains one of the most profound challenges in theoretical physics. It calls into question our understanding of the deep interplay between gravity, quantum mechanics, and the structure of spacetime. While much progress has been made in higher-dimensional settings, lower-dimensional models continue to provide invaluable insights thanks to their tractability and their ability to capture essential features of black hole dynamics, such as apparent and event horizons, curvature singularities, Hawking radiation, and the potential breakdown of unitary quantum evolution. Among these, the two-dimensional Callan-Giddings-Harvey-Strominger (CGHS) model \cite{CGHS} has proven especially fruitful  for probing both classical and quantum aspects of black holes, including evaporation and the fate of the curvature singularity.

In this paper, we revisit the CGHS model of two-dimensional dilaton gravity conformally coupled to $N$ scalar fields. We aim to investigate a mechanism for the quantum resolution of spacelike curvature singularities and the restoration of unitarity in the dynamical ``evaporation'' process. Specifically, we examine how a total negative central charge in the effective one-loop theory alters the standard semiclassical picture, which is typically constructed in the large-$N$ limit and therefore characterized by a positive central charge. This line of inquiry was originally initiated in Refs.~\cite{PSS22, PSS23, PSS23b} for the Russo--Susskind--Thorlacius (RST) one-loop corrected version of the CGHS model~\cite{RST}, and further developed in Ref.~\cite{dRMN25} for the case of spherically reduced Einstein gravity. A key result emerging from both models is the following: the static, radiationless, and asymptotically flat solutions of the conventional semiclassical equations (with positive central charge) transform the classical black hole geometry into a horizonless spacetime featuring a curvature singularity. Crucially, in these static models, reversing the sign of the central charge (i.e., making it negative) eliminates this singularity, yielding a backreacted geometry that is still horizonless  and remains asymptotically flat. 

In conventional treatments that incorporate backreaction during gravitational collapse induced by conformal matter, a spacelike singularity forms and serves as a final boundary for infalling information. This viewpoint, grounded in Hawking's original argument, results in non-unitary evolution and the apparent loss of information within the semiclassical CGHS model, as analytically described in~\cite{RST}. In contrast, it is found analytically that, under suitable conditions, a dominant negative central charge completely eliminates the classical singularity in the dynamical formation of a CGHS black hole. This result was anticipated in~\cite{Strominger92}, where it was shown that, in the vicinity of an incoming matter shock wave, the backreacted geometry is non-singular. We confirm here that the geometry is, in fact, non-singular everywhere. We reach this conclusion by constructing an analytically solvable semiclassical model in which the Faddeev--Popov ghost fields associated with the two-dimensional reparametrization invariance are prevented from contributing to Hawking radiation, as required in~\cite{Strominger92}. This solvable model turns out to be mathematically equivalent to the RST model with $N$ replaced by a negative quantity, as first analysed in~\cite{PSS22, PSS23, PSS23b}. Our construction allows us to compute explicitly the physical radiation approaching the asymptotically flat regions within the apparent horizon, at both left and right null infinities. We identify regions in which the outgoing radiation flux becomes negative and is correlated with the Hawking radiation emitted outside the horizon, paving the way to a unitary description of the dynamical evolution.

The paper is organized as follows. In Section~\ref{section2}, we begin by reviewing the classical aspects of the CGHS model and its thermodynamical behavior, including the Hawking flux.  In Section~\ref{sec:oneloop-extension}, we introduce our one-loop extension of the classical CGHS model. The one-loop effective action consists of three terms, each playing a fundamental physical role. 
One term corresponds to the positive central charge of the $N$ scalar fields and is represented by the standard non-local Polyakov action \cite{Polyakov81}. Another term is primarily associated with the fluctuations of the Faddeev-Popov (FP) ghosts arising from the FP determinant introduced when fixing the gauge for the reparametrization invariance of the two-dimensional gravitational theory \cite{Polyakov81}. The central charge of the $ b-c$ ghost system is negative, corresponding to a non-unitary theory. Therefore, it is crucial to prevent these ghost fields from appearing as quanta emitted by Hawking radiation.
This is achieved through Strominger's mechanism \cite{Strominger92}, which constructs the induced Polyakov action with respect to an auxiliary metric that remains flat and decoupled from the collapse described by the physical metric. To ensure the consistency of this mechanism, the auxiliary metric must remain flat at the one-loop level. This condition can be satisfied by adding appropriate local counterterms. Moreover, a specific counterterm is uniquely selected by requiring that two-dimensional Minkowski space remains an exact solution of the backreacted, one-loop theory. In Section~\ref{sectionLargeN}, we present the semiclassical description of the evaporation process in the large $N$ limit, where the contribution of ghosts and the gravity sector can be neglected. An apparent horizon forms alongside a spacelike singularity, and the two meet at a finite point, which marks the endpoint of the evaporation process. At leading order, the process appears consistent with energy conservation. However, a closer examination reveals a violation of energy conservation due to the presence of the spacelike singularity.

These sections provide essential ingredients for understanding Section~\ref{sec:backreacted solutions negative C}, which explores a regime opposite to the conventional large $N$ limit. Here, we investigate the quantum dynamics of black hole formation in a scenario where the total central charge is dominated by ghosts rather than matter, and hence becomes negative.\footnote{A systematic line of research in this direction was initiated in \cite{PSS22, PSS23, PSS23b}; however, our construction of the relevant model and its physical interpretation differ from those proposed there. In particular, we employ Strominger's mechanism \cite{Strominger92} to suppress the Hawking emission of ghost fields.} This regime naturally incorporates effects from the quantization of the metric, as ghosts arise when accounting for diffeomorphism invariance of the gravity theory within the functional integral formalism. Consequently, we expect our results to align with those from quantum geometry, which, in various simplified models such as the two-dimensional CGHS model \cite{ATV08, APR11}, strongly suggests that spacelike singularities are resolved.
Such a resolution fundamentally reshapes our perspective on the unitarity problem. Early discussions typically assumed that the future boundary of spacetime consists not only of future null infinity but also of a segment of the classical singularity. This is what happens in the large $N
$ approximation described in Section~\ref{sectionLargeN}. Under this assumption, a portion of the initial state inevitably falls into the singularity, from which it follows that the outgoing state at future null infinity cannot preserve all the correlations contained in the incoming state at past null infinity. However, if the singularity is resolved, this potential obstacle to preserving the purity of the incoming state is eliminated. We also analyze the energy balance of the model, finding that negative energy fluxes evaluated at null surfaces approaching left and right infinities, respectively, cannot account for the standard, positive-energy, quasi-thermal Hawking flux. Finally, Section~\ref{secConclusions} summarizes our conclusions and  outlines a possible extension of the model aimed at recovering energy conservation while preserving unitarity.

\section{The classical CGHS model} \label{section2}

In this section we will briefly review the classical aspects of the two-dimensional theory of gravity coupled to a dilaton field $\phi$ and with a matter sector given by a set of $N$ massless scalar fields $f_i$, as first introduced in \cite{CGHS}.  The action is given by
\bea \label{classicalCGHS}
  S_{\text{CGHS}}={1\over2\pi}\int d^2
 x\sqrt{-g}[e^{-2\phi}\left(R+4\left(\nabla\phi\right
 )^2+4\lambda^2\right)
 -{1\over2}\sum_{i=1}^N\left(\nabla f_i\right)^2]
  \>. 
 \eea
The classical field equations which follow from the above action are
\begin{equation}
e^{-2 \phi}\left[R+4 \lambda^2+4 \nabla^2 \phi-4(\nabla \phi)^2\right]=0,
\end{equation}

\begin{equation}
2 e^{-2 \phi}\left[\nabla_\mu \nabla_\nu \phi+g_{\mu \nu}\left((\nabla \phi)^2-\nabla^2 \phi-\lambda^2\right)\right]=T_{\mu\nu}^f,
\end{equation}
where $T_{\mu\nu}^f$ is the matter stress-energy tensor 
\begin{equation}
T_{\mu \nu}^f=\frac{1}{2} \sum_{i=1}^N\left[\nabla_\mu f_i \nabla_\nu f_i-\frac{1}{2}g_{\mu\nu}\left(\nabla f_i\right)^2\right] \ . 
\end{equation}

In conformal gauge $ds^2 = -e^{2\rho} dx^+dx^-$ the above field equations become
\begin{equation}
\begin{aligned}
& e^{-2(\phi+\rho)}\left[-4 \partial_{+} \partial_{-} \phi+4 \partial_{+} \phi \partial_{-} \phi+2 \partial_{+} \partial_{-} \rho+\lambda^2 e^{2 \rho}\right]=0 \\
& e^{-2 \phi}\left[2 \partial_{+} \partial_{-} \phi-4 \partial_{+} \phi \partial_{-} \phi-\lambda^2 e^{2 \rho}\right]=0 \ , 
\end{aligned}
\end{equation}
together with the constraint equations involving the matter stress-energy tensor
\bea
e^{-2 \phi}\left(-4 \partial_{ \pm} \rho \partial_{ \pm} \phi+2 \partial_{ \pm}^2 \phi\right)= T_{\pm\pm}^f \equiv \frac{1}{2} (\partial_{\pm}f)^2\, .
\eea
Combining the first two equations we infer that $(\rho - \phi)$ obeys a free field equation
\be \label{ffeq} \partial_+\partial_-(\rho -\phi)=0 \ . \ee
In covariant language this condition can be easily re-expressed introducing the re-scaled metric $\hat g_{\mu\nu}= e^{-2\phi}g_{\mu\nu}$. The new metric turns out to be a flat metric $R(\hat g)=0$ when the field equations are satisfied. This arises as a consequence of a hidden  symmetry  of the classical theory 
    \be  \delta g_{ab}=2e^{2\phi} g_{ab}  \ \ \ \ \ \ \delta \phi=e^{2\phi} \ , \ee
 implying the above free field equation. The general solution for \eqref{ffeq} is a sum of two chiral functions $\omega(x^+)+ \omega(x^-)$. A coordinate transformation within the conformal gauge change the form of those functions, and one can use this gauge ambiguity to fix coordinates by imposing $\omega(x^+)=0$ and $\omega(x^-)=0$, and hence $\rho-\phi=0$. This condition is called the Kruskal gauge. It is a very convenient gauge to solve the field equations, which now become
 \be \partial_+\partial_- e^{-2\phi} = -\lambda^2 \ , \ee
 and 
 \be e^{-2 \phi}\left(-4 (\partial_{ \pm} \phi)^2+ 2\partial_{ \pm}^2 \phi\right)\equiv \partial_{\pm}^2 e^{-2\phi}= T_{\pm\pm}^f
 \ . \ee
 In the absence of matter fields the general solution is
 \be e^{-2\rho}=e^{-2\phi}= -\lambda^2 x^+x^- + \frac{M}{\lambda}\, . \ee
 
 It describes a black hole with future and past event horizons at $x^+=0$ and $x^-=0$, respectively.  A curvature singularity ($R=8e^{-2\rho}\partial_+\partial_-\rho \to \infty$)  is hidden in the interior along the spacelike curve $-\lambda^3 x^+ x^- +M=0$. The configuration with $M=0$ represents the two-dimensional Minkowski space. It is  also called the linear dilaton vacuum.

\subsection{Black hole formation}
  
A black hole can be easily formed by sending a null shell  of matter, with stress-energy tensor 
\be \label{nullshell}T_{++}^f = \frac{m}{\lambda x^+_0}\delta(x^+-x^+_0) \ , \ee
and energy $m$. The dynamical solution is 
\bea\label{dynamicalsol}
e^{-2\rho}=e^{-2\phi}
=-\lambda^2x^+x^- 
-{m\over\lambda x^+_0}\left(x^+-x^+_0
\right)\theta\left(x^+-x^+_0\right)
\>,\label{eq:cdynamical}
\eea
and describes the formation of a black hole of mass $m$. Note that, since $e^{-2\phi}$ is a positive function,  the spacetime is defined in the region $0<x^+<+\infty$ and $-\infty < x^- <0$.

\begin{figure}[h]
\includegraphics[angle=0, width=100mm]{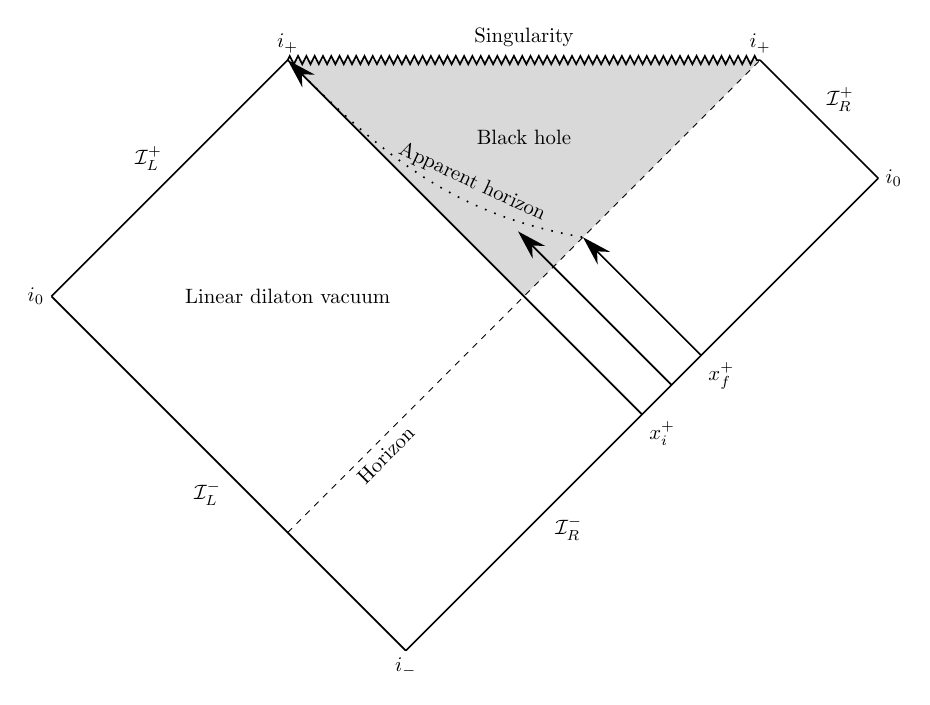}
\caption{\small Penrose diagram illustrating gravitational collapse due to a continuous influx of  matter. The ingoing flux starts at some time $x^+_i$ and vanishes after the time $x^+_f$. An apparent horizon forms, whose location is sensitive to the specific details of $T_{++}^f$. In contrast, the event horizon depends solely on the integrated quantity $P(x^+_f)$. The resulting curvature singularity is spacelike and lies hidden behind the event horizon, consistent with the cosmic censorship conjecture.} 
\label{cghsBLACKHOLE}
\end{figure}

For a general form of $T_{++}^f(x^+)$, with support in the interval $[x^+_i, x^+_f]$, the solution is given by
\bea
e^{-2\rho}=e^{-2\phi}
=-\lambda^2x^+(x^- + P(x^+)/\lambda^2) 
+{m(x^+)\over\lambda}
\>,\label{eq:cdynamical1}
\eea
where 
\be m(x^+)= \lambda \int_{x^+_i}^{x^+} dy^+\,  y^+T_{++}(y^+) \ee
and 
\be P(x^+)= \int_{x^+_i}^{x^+} dy^+\,  T_{++}(y^+) \ . \ee
The event horizon is given by the null line
\be x^-= x^-_h \ , \ee
where $x_h^-=-\frac{P(x^+_f)}{\lambda^2}$, and the apparent horizon, defined by the condition $\partial_+ e^{-2\phi}=0$, is given, for $x^+\ge x^+_i$,  by the curve
\be x^-= -\frac{P(x^+)}{\lambda^2} \ . \ee
For $x^+ \ge x^+_f$ it coincides with the event horizon. The emerging curvature singularity is hidden beyond the horizon and it is given by the spacelike curve
\be x^-= -\frac{P(x^+)}{\lambda^2} + \frac{m(x^+)}{\lambda^3 x^+} \ . \ee
These features are represented in Fig \ref{cghsBLACKHOLE}. 

In the asymptotic null infinity $x^+\to +\infty$, one gets $R= 8e^{-2\rho}\partial_+\partial_- \rho \to 0$. In this region the metric is flat and behaves  as
\be 
ds^2 = -e^{2\rho}dx^+dx^- \sim -\frac{dx^+}{\lambda x^+}\frac{dx^-}{-\lambda (x^- - x^-_h)} \ . 
\ee
One can compare the form of this asymptotic solution with the solution before the incoming matter ($x^+ < x_i^+$)

\be 
ds^2 = -\frac{dx^+}{\lambda x^+}\frac{dx^-}{-\lambda x^-} = -d\sigma^+ d\sigma^- \ , 
\ee
where we have defined 
\bea \label{sigma+}\lambda\sigma^+&=&\log\lambda x^+ \ , \\
\lambda\sigma^-&=&-\log\left(-
\lambda x^-\right) \label{sigma-} \ . \eea
The coordinates $(\sigma^+, \sigma^-)$ define a set of null Minkowskian  coordinates before the collapse.  Note that any other possibility is obtained by a Lorentz transformation, $\sigma^+ \to \gamma \sigma^+$, $\sigma^- \to \gamma^{-1}\sigma^-$, where $\gamma$ is a positive constant.  Those defined in \eqref{sigma+} are fixed by the condition 
\be 
\int_{-\infty}^{+\infty} d\sigma^+\, T^f_{\sigma^+\sigma^+}= m(x^+_f) 
\ . 
\ee 
Going back to the asymptotic null region, and demanding now that 
\be 
\label{dx-dhatsigma-}
\frac{dx^-}{-\lambda (x^- -x^-_h)}= d\hat \sigma^- \ , 
\ee
we can identify a null set  of conformally flat coordinates $(\hat\sigma^+=\sigma^+, \hat \sigma^-)$ at $x^+\to +\infty$. Up to trivial translations, we easily get 
\bea 
\label{hatsigma+}
\lambda\hat\sigma^+&=&\log\lambda x^+ \ , \\
\lambda\hat\sigma^-&=&-\log(-\lambda(x^--x^-_h)) \label{hatsigma-} \ . 
\eea

\subsection{Hawking radiation}\label{Hawkingradiation}

Given the classical background geometry, the outgoing Hawking flux can be obtained from the trace anomaly
\be 
\label{traceanomaly}
\langle T \rangle = \frac{C}{24} R\ , 
\ee
where $C$ is the central charge of the underlying conformal field theory of the matter sector. For $N$ massless scalar fields $C=N$. The  conformal anomaly in two dimensions and the conservation law $\nabla^\mu \langle T^{f}_{\mu\nu} \rangle =0$ constrain the form of the stress-energy tensor 
\be 
\label{mastereq}
\langle T_{\pm\pm}^f\rangle =-{N\over12}\left[\left(\partial_{\pm} \rho\right)^2-\partial^2_{\pm} \rho + t_{\pm}\right] \ . 
\ee
The integration functions $t_{\pm}$ are chiral functions that characterize the vacuum state implicit in the above vacuum expectation values (see, for example, \cite{FN05}).\footnote{The fundamental expression~\eqref{mastereq} can be obtained directly  from the anomalous transformation law of the normal-ordered  stress-energy tensor in two dimensions, $\langle : T_{\pm\pm} : \rangle$, together with the equivalence principle~\cite{FN05}. The  functions $t_{\pm}$ are then identified via $\langle : T_{\pm\pm} :  \rangle = -\frac{N}{12} t_{\pm}$. From this, the trace anomaly  follows readily as a consequence of~\eqref{mastereq}.} The functions $t_{\pm}$ transform, under a conformal change of coordinates, according to  
\be 
\label{Schderivative}
t_{\pm}(y^{\pm})= \left(\frac{dz^{\pm}}{dy^{\pm}}\right)^2t_{\pm}(z^{\pm}) + \frac{1}{2}\{z^{\pm}, y^{\pm}\} \ ,  
\ee
where $\{z^{\pm}, y^{\pm}\}$ is the Schwarzian derivative
\be 
\{z^{\pm}, y^{\pm}\}\equiv \frac{d^3z^{\pm}}{d y^{\pm 3}} \left(\frac{dz^{\pm}}{d y^{\pm}}\right)^{-1} -\frac{3}{2}\left(\frac{d^2z^{\pm}}{dy^{\pm 2}}\right)^2 \left(\frac{dz^{\pm}}{dy^{\pm}}\right)^{-2}\ . 
\ee
When $x^+<x^+_i$ the spacetime metric is given by
\be 
ds^2 = -d\sigma^+d \sigma^- \ . 
\ee
The quantum vacuum coincides with the Minkowskian vacuum at $x^-<x^-_i$ and then \be 
\label{eq:minkowskivacuum0} 
\langle  T^f_{\sigma^{\pm}\sigma^{\pm}}\rangle = t_{\pm}(\sigma^{\pm}) =0 \ . 
\ee
In coordinates $x^{\pm}$ the functions $t_{\pm}(x^{\pm})$ take, according to \eqref{Schderivative}, the form
\be 
t_{\pm}(x^{\pm})=\frac{1}{(2x^{\pm})^2} \ . 
\ee
The Hawking flux can be then evaluated from the form of $\langle T^f_{x^-x^-}\rangle $ at $x^+ \to +\infty$. From \eqref{mastereq} one gets 
\be 
\label{fluxx-}
 \langle T^f_{x^-x^-}\rangle_{\mathcal{I}^+_R} = \frac{N}{48}\left[\frac{1}{(x^- - x^-_{h})^2} - \frac{1}{(x^-)^2}\right]
\ . 
\ee
Therefore, in  asymptotically flat coordinates, one obtains
\bea 
\label{scriIflux}
\langle T^f_{\hat \sigma^-\hat\sigma^-}\rangle_{\mathcal{I}^+_R} &=& \left ( \frac{dx^-}{d\hat \sigma^-}\right )^2_{\mathcal{I}^+_R} \langle T^f_{x^-x^-}\rangle_{\mathcal{I}^+_R}\nonumber \\
&=&\frac{N \lambda^2}{48}\left[1-\left(1+\lambda^{-1} P(x^+_f) \, e^{\lambda \hat \sigma^{-}} \right)^{-2}\right]
\eea

The emission flux vanishes exponentially at early times $\hat \sigma^- \to -\infty$ and, after a transitory period, goes to the thermal constant $N\lambda^2/48$ at late times $\hat \sigma^-\to +\infty$. The Hawking temperature is determined by this constant and it is given by $T_H= \lambda/2\pi$. Note that of all the classical information imprinted in $T^f_{++}(x^+)$, only the single momentum $P(x^+_f)$ is encoded in the Hawking flux. For the  shock wave \eqref{nullshell}, we have $P(x^+_f)= m/\lambda x^+_0$, with  $x^+_i=x^+_0=x^+_f$. 

For the remainder of this work, and for the sake of simplicity, we restrict our analysis to the case of such an incoming shock wave.

\subsubsection{Outgoing flux along the null surface $x^+=x^{+}_r$}\label{outgoingflux}

For future purposes, it is convenient to redo the calculation by first evaluating the outgoing flux along the null surface $x^+=x^+_r$, where $x^+_r$ is a reference point. We take the limit $x^+_r\to +\infty$ at the end of the calculation. The flux $\langle T^f_{x^-x^-}\rangle (x^-, x^+_r)$ is given by
\bea 
\langle T^f_{x^-x^-}\rangle (x^-, x^+_r)&=& -{N\over12}\left[\left(\partial_{-} \rho\right)^2-\partial^2_{-} \rho + t_{-}\right] \nonumber \\
[0.5em]
&=& \frac{N}{48} \left[\frac{(\lambda^2 x^+_r)^2}{(-\lambda^2x^+_r(x^--x^-_h) + \frac{m}{\lambda})^2} -\frac{1}{(x^-)^2}\right]
\ . 
\eea
Note that, in the limit $x^+_r\to +\infty$, we easily reproduce \eqref{fluxx-}. Physically, it is more appropriate to consider the outgoing flux as measured in terms of an affine parameter $\hat \sigma^-$ along the geodesic $x^+=x^+_r$. [In the limit $x^+_r\to +\infty$, this affine parameter coincides with \eqref{hatsigma-} and for this reason we maintain the notation $\hat\sigma^-$ for the affine parameter even for finite $x^+_r$]. Therefore the relevant flux is given by the quantity 
\be 
\langle T^f_{\hat \sigma^-\hat \sigma^-}\rangle \equiv \left ( \frac{dx^-}{d\hat \sigma^-}\right )^2_{x^+=x^+_r}\langle T^f_{x^-x^-}\rangle  (x^-, x^+_r)\ . 
\ee
The value of  $( \frac{dx^-}{d\hat \sigma^-})_{x^+=x^+_r}$ 
 can be easily obtained from the geodesic equation 
 \begin{equation}
 \frac{d^2x^-}{d\hat \sigma^{-2}} +\Gamma_{--}^- \frac{dx^-}{d\hat \sigma^{-}}\frac{dx^-}{d\hat \sigma^{-}}=0 \ . 
\end{equation}
 Since $\Gamma_{--}^- =2\partial_-\rho$ we can easily find the general solution along the null  geodesic $x^+=x^+_r$ 
 \begin{equation}
\frac{dx^-}{d\hat \sigma^{-}} = e^{-2\rho(x^-,\, x^+_r) +A(x^+_r)}
 \end{equation}
 The functional dependence of the integration constant  $A(x^+_r)$  can be fixed by boundary conditions. We can determine it by assuming that, when $x^+_r < x^+_0$ the solution is the two-dimensional Minkowski  space $e^{2\rho}= (-\lambda^2 x^+x^-)^{-1}$.  We then have 
 \be 
 \frac{dx^-}{d\hat \sigma^{-}}= -\lambda x^-= e^{-2 (\frac{1}{2}\log (-\lambda^2 x_r^+ x^-)^{-1})+A(x^+_r)}= -\lambda^2 x^-x^+_r e^{A(x^+_r)} 
 \ee 
Therefore, $A(x^+_r)= -\log \lambda x_r^+$ and get the general expression
\begin{equation}
\frac{dx^-}{d\hat \sigma^{-}} = \frac{1}{\lambda x_r^+}e^{-2\rho(x^-,\, x^+_r)}\, .
 \end{equation}
The flux is then given by the expression
\bea 
\langle T^f_{x^-x^-}\rangle \left ( \frac{dx^-}{d\hat \sigma^-}\right )^2_{x^+=x^+_r}&=& \frac{N}{48} \left[\frac{(\lambda^2 x^+_r)^2}{(-\lambda^2x^+_r(x^--x^-_h) + \frac{m}{\lambda})^2} -\frac{1}{(x^-)^2}\right]\nonumber \\
&\qquad &  \,\,\, \times\left[-\lambda(x^--x^-_h) +\frac{m}{\lambda^2 x^+_r}\right]^2 
\eea
In the limit $x^+_r\to +\infty$ we recover the expression \eqref{scriIflux}. 

The total radiated  energy crossing the null surface $x^+=x^+_r$ in the region outside the  horizon is given by the integral
\bea
E_{\text{rad}}    (x^+_r)&=& \int_{-\infty}^{\hat \sigma^-_h} d\hat \sigma^-\langle T^f_{x^-x^-}\rangle \left ( \frac{dx^-}{d\hat \sigma^-}\right )^2_{x^+=x^+_r} = \int_{-\infty}^{x^-_h} dx^-\langle T^f_{x^-x^-}\rangle \left ( \frac{dx^-}{d\hat \sigma^-}\right )_{x^+=x^+_r}  \nonumber \\
[0.7em]
&=&\frac{N}{48}\left[\lambda + \frac{m}{\lambda^{2} x_r^{+} x_h^{-}}
   + \lambda \ln\!\left|\frac{\lambda^{3} x_r^{+} x_h^{-}}{m}\right|\right]\ 
\eea
As expected, $E_{\text{rad}}(x^+_r)$ diverges as $x^+_r \to +\infty$. This behavior reflects the fact that, in the classical theory, the black hole is eternal and therefore radiates indefinitely.

We finish this section by recalling two fundamental, closely related problems intrinsic to the fixed-background approximation adopted so far and to the presence of a curvature singularity hidden behind the event horizon:

\begin{enumerate}

\item Correlations are lost: (i) the $S$-matrix from
$\mathcal{I}^-_L$ to $\mathcal{I}^+_R$ is non-unitary; (ii) the
$S$-matrix from $\mathcal{I}^-_R$ to $\mathcal{I}^+_L$ is also
non-unitary. This is a direct consequence of the singularity, which
abruptly truncates the spacetime.

\item Without backreaction, energy conservation cannot be
maintained, as already evident from the divergence of
$E_{\text{rad}}$.

\end{enumerate}
We will see in the following sections how the one-loop extension of the classical theory deals with these issues.

\section{One-loop extension of the classical theory}
\label{sec:oneloop-extension}

It is useful to consider the quantum theory of our classical model \eqref{classicalCGHS} in terms of the functional integral
\be \mathcal{Z} = \int \mathcal{D}(g, \phi, f_i)\,e^{iS_{\text{CGHS}}} \ . \ee
This expression can be improved by introducing a gauge fixing to avoid the overcounting of metrics due to the underlying general covariance. We choose to work in the conformal gauge. Therefore, the functional integral over metrics is replaced by an integral over the conformal factor $\rho$, where $ds^2=-e^{2\rho}dx^+dx^-$, and the anticommuting $(b,c)$ Faddeev-Popov ghosts  \cite{Polyakov81, Strominger92}
\be 
\label{cghsFP}
\mathcal{Z} = \int \mathcal{D}(\rho, \phi, b, c, f_i)\, e^{i(S_{\text{CGHS}}+ S_{ghosts})} \ . 
\ee
The above functional integral can be evaluated in the one-loop approximation, which means to integrate out the quadratic fluctuations around a background configuration. This integration is exact for the matter and ghost sectors since they are already quadratic. The functional integration over the matter fields $f_i$ produces the well-known Polyakov-Liouville (PL) action, whose effective Lagrangian, in conformal gauge, is $\frac{N}{12\pi} \partial_+\rho \partial_-\rho$. The covariant form of the effective Polyakov-Liouville action is 
\be 
\label{Gamma++}\Gamma_+\left[g_{\mu\nu}\right] =  -\frac{N}{96\pi}\int d^2x \sqrt{-g}\,R\Box^{-1} R\ . 
\ee
It is important to note that, in performing this integration, a covariant definition of the measure $\mathcal{D}(f_i)$ has been adopted with respect to the metric $ds^2 = -e^{2\rho} dx^+ dx^-$
\begin{equation}\label{fsector}
\mathcal{D}(f_i)= \mathcal{D}_0(f_i) e^{-\frac{iN}{12\pi}\int d^2x \, \partial_+\rho  \partial_-\rho} \ , 
\end{equation}
where $\mathcal{D}_0(f_i)$ denotes the measure defined with respect to the flat metric, corresponding to $\rho = 0$.
The action \eqref{Gamma++} is non-local, reflecting the freedom we have to choose the quantum state of the matter fields. The stress-tensor obtained from $\Gamma_+\left[g_{\mu\nu}\right]$ reproduces \eqref{mastereq}. The functions $t_{\pm}$ reflect the non locality of the Polyakov-Liouville effective action. 

The integration of the ghost fields produces a similar effective action, with the replacement of the positive number $N$ by the negative number $-26$ (see, for example,  \cite{Polyakov81, GSW87, Tong09}). Furthermore, the integration of the fluctuations of the fields $\rho$ and $\phi$  shifts $-26$ to $-24$ . Therefore, the overall coefficient  of the Polyakov-Liouville action is proportional to $N-24$ \cite{Strominger92}. In the large $N$ limit one can ignore the contributions of the ghosts and the fluctuations of the fields $\rho$ and $\phi$. In any case, and for $N< \infty$, a major difficulty of the above line of reasoning is that a reevaluation of the Hawking flux produces a contribution proportional to $N-24$, instead of  being proportional to the number of degrees of freedom of matter $N$, as in \eqref{scriIflux}.
This would imply the spontaneous emission of ghosts in addition to the conventional Hawking emission for matter fields.
One can avoid this implication following the mechanism proposed by Strominger in \cite{Strominger92}. The basic idea is to take advantage of the fact that the auxiliary metric $\hat g_{\mu\nu}=e^{-2\phi}g_{\mu\nu}$ is a flat metric in the classical theory. [We advance here that the auxiliary metric will remain flat in the one-loop extension of the theory]. 
The mechanism assumes that the ghost fields   are coupled to  $\hat g_{\mu\nu}$, instead of the physical metric $g_{\mu\nu}$. 
 This assumption is equivalent to take the measure for the ghost sector as
\begin{equation}
\mathcal{D}(b,c)= \mathcal{D}_0(b,c )  \ , 
\end{equation}
instead of \eqref{fsector}.
This reasoning is justified by the absence of any fundamental requirement to employ the same metric in constructing the functional integral for the matter fields and for the ghost sector \cite{Harvey-Strominger92}.
Since the metric $\hat g_{\mu\nu}$ is an on-shell flat metric, due to the global symmetry of the  CGHS action, the emission of ghost quanta in the process of black hole formation is automatically blocked, as one should expect for  ghost fields \cite{Strominger92, Harvey-Strominger92, Strominger, Liberati}. In this way, the ghost fields can play a non-trivial role without contributing to physical excitations. 
The argument can also be extended  to the remaining degrees of freedom in conformal gauge, namely, the $\phi$ and $\rho$ fields. Since they are not propagating degrees of freedom, their contribution to the effective action should be defined in such a way that no artificial Hawking radiation is associated to those fields. This can also be guaranteed by using the metric $\hat g_{\mu\nu}$ in the definition of the measure for those fields in the functional integral \cite{Strominger92}.  The contribution to the effective action of those (unphysical) fields should be of the form
\be \label{Gamma-}\Gamma_-\left[\hat g_{\mu\nu}\right] =  -\frac{C_-}{96\pi}\int d^2x \sqrt{-\hat g}\, R(\hat g)\Box^{-1}(\hat g) R(\hat g)\ , \ee
where $C_-=-24$ is the total (negative) central charge coming from ghosts and $(\rho, \phi)$ fields.

 A theory with a negative central charge is non-unitary, but it is not an issue here since we are not allowing physical excitations (Hawking radiation) associated to those fields.  This is implemented by the use of the metric $\hat g$ in the effective Liouville-Polyakov action. The contribution of \eqref{Gamma-} to the  stress-energy tensor is
\be 
\label{mastereq-}
\langle T_{\pm\pm}\rangle =-{C_-\over12}\left[\left(\partial_{\pm} \hat \rho\right)^2-\partial^2_{\pm} \hat \rho + \hat t_{\pm}\right] \ . 
\ee
 Since $\hat g$ is assumed a flat metric everywhere, one can enforce the gauge $d\hat s^2= -dx^+dx^-$. Moreover, recalling that  the coordinates $(x^+, x^-)$ are naturally restricted to the region $0< x^+<+\infty$, and  $0>x^->-\infty$ (see the comment after \eqref{dynamicalsol}), we can identify the geometry of this region in the metric $\hat g_{\mu\nu}$ as the right Rindler wedge of Minkowski space with $d\hat s^2 =-dx^+dx^-$. 
The natural and unrestricted coordinates for the right Rindler wedge are indeed $\sigma^{\pm}$, defined as $\lambda x^{\pm}=\pm  e^{\pm\lambda \sigma^{\pm}}$. The metric takes then a stationary form $d\hat s^2= - e^{\lambda(\sigma^+-\sigma^-)}d\sigma^+d\sigma^-$ everywhere. The metric $\hat g$ is insensitive to the matter fields and no Hawking radiation can be induced by incoming matter fields. Furthermore, the natural vacuum state of the ghosts fields is a Rindler-type vacuum $|0_R\rangle$, which is characterized by the positive frequency modes $e^{-i\omega \sigma^+}$ and $e^{-i\omega \sigma^-}$, or, equivalently, the boundary conditions $\hat t_{\pm}(\sigma^{\pm})=0$ [Recall also Eq.\eqref{eq:minkowskivacuum0} of the previous section to characterize the vacuum of the matter fields]. This implies
\be \hat t_{\pm}= \frac{1}{(2x^{\pm})^2} \ . \ee

The full action to describe our one-loop theory, including backreaction, is given by $ S=\, S_{\text{CGHS}}[g, \phi, f] + \Gamma_-[\hat g] + \Gamma_+[g]$. However, this is not fully consistent.  The  introduction  of the metric $\hat g$ for constructing the effective action associated to  $\Gamma_-[\hat g]$ is to block the spontaneous emission of unphysical excitations. But this is only guaranteed if the metric $\hat g$ is a flat metric also in the one-loop theory. This is the case only if $N=0$, otherwise the underlying classical symmetry is broken. For $C_+\equiv N >0$ one can fix this problem by introducing local counterterms such that the global symmetry is restored and the metric $\hat g$ continues to be a flat metric in the one-loop theory.  A one-parameter family of local counterterms doing this job is given by \cite{CN96}
\be 
S_{local} [g, \phi] = \frac{C_+}{24\pi} \int d^2 x \sqrt{-g} \, [(1-2a) (\nabla \phi)^2 + (a-1)\phi R] \ , \ee
where $a$ is a real parameter. Therefore, our one-loop theory is defined by 
\be \label{CGHSoneloop} S=\, S_{\text{CGHS}}[g, \phi, f] + \Gamma_-[\hat g] + \Gamma_+[g] + S_{local}[g, \phi] \ . 
\ee

The  ambiguity in the parameter $a$ can be fixed by imposing a simple condition. We demand the two-dimensional Minkowski space to be an exact solution of the one-loop theory. As we will see later, this fixes $a$ as follows 
\be \label{a=} a=a_0\equiv \frac{C_- + C_+}{2C_+} \ . \ee

Note that, before the incoming matter is switched on, the contribution of $\Gamma_-[\hat g_{\mu\nu}]$ to the stress-energy tensor \eqref{mastereq-} is canceled by that of $S_{local}[g, \phi]$ for the choice \eqref{a=}. Furthermore, neither $\Gamma_-[\hat g_{\mu\nu}]$ nor $S_{local}[g, \phi]$ contribute to the ingoing and outgoing physical radiation.

After some manipulations, the semiclassical equations derived from the effective action \eqref{CGHSoneloop} can be conveniently written as
\be
\partial_+\partial_-\left(e^{-2\phi}+{C_+\over12}a\phi\right)
+\lambda^2e^{2\left(\rho-\phi\right)}=0\>,\label{xxix}
\ee
\be
\partial_+\partial_-\left(\rho-\phi\right)=0\>,\label{xxx}
\ee
\bea
\partial^2_{\pm}\left(e^{-2\phi}+{C_+\over12}a\phi\right)
-2\partial_{\pm}\left(\rho-\phi\right)\partial_{\pm}\left(
e^{-2\phi}+{C_+\over12}a\phi\right) & \nonumber \\
+T^f_{\pm\pm} -{C_+\over12}\left[\left(\partial_{\pm}\left(\rho-\phi\right)
\right)^2-\partial^2_{\pm}\left(\rho-\phi\right) + t_{\pm}\right] & \nonumber \\
-{C_-\over12}\left[\left(\partial_{\pm}\left(\rho-\phi\right)
\right)^2-\partial^2_{\pm}\left(\rho-\phi\right) + \hat t_{\pm}\right] & =0\>.\label{xxxi}
\eea
The chiral functions $t_{\pm}$ and $\hat t_{\pm}$ are integration  functions determined by the specific quantum states of the matter  and ghost-gravity sectors, respectively. Since we impose that the initial configuration is the two-dimensional Minkowski space for the physical metric $g$, these integration functions are given, in the gauge $\rho = \phi$, by 
\be
\label{tfixed} t_{\pm}(x^{\pm}) = \frac{1}{(2x^{\pm})^2}= \hat t_{\pm}(x^{\pm}) \ . 
\ee
The prescription  $t_{\pm}=1/(2x^{\pm})^2$ also selects a vacuum in which the quantum incoming flux vanishes at past null infinity, both during and after the gravitational collapse. \\

\section{The conventional semiclassical large $N$ limit} \label{sectionLargeN}

The full quantum theory defined by \eqref{cghsFP} is rather involved. The theory simplifies at the  one-loop semiclassical level, as defined by the effective action \eqref{CGHSoneloop}. This approach is refined by considering the large $N$ approximation, which also assumes the product $Ne^{2\phi}$ to be held fixed. Although the primary goal of this paper is to  investigate the unexplored regime of finite $N$ for a negative total central charge $C_++C_- \equiv N -24 <0$, it is nevertheless useful, both for our purposes and to provide a broader  perspective of our approach,  to briefly describe in this section the conventional semiclassical picture.

In this situation we can ignore the contribution $\Gamma_-[\hat g]$ and work with the remaining action. For $C_-=0$ the parameter $a$ becomes $a=1/2$ and the semiclassical theory is governed by the action

\be 
\label{oneloop}
S_{\text{RST}} [g, \phi, f_i]= S_{\text{CGHS}}   -\frac{N}{96\pi}\int d^2x \sqrt{-g}\, R\Box^{-1} R - \frac{N}{48\pi} \int d^2 x \sqrt{-g}\,  \phi R \ , \ee
which turns out to be the semiclassical  RST model \cite{RST}.\footnote{It is also a common practice in the literature to regard $S_{local}[g, \phi; a=1/2]$ as an extra contribution to the classical model, and consider $S_{\text{CGHS}}[g, \phi, f_i]+S_{local}[g, \phi; a=1/2]$ as the classical action \cite{Strominger2020}.} This semiclassical model allows us to describe analytically the  formation and  evaporation of the two-dimensional CGHS black hole. 

For our purposes it is convenient to analyze the semiclassical theory for generic parameter $a$. The semiclassical equations are  given by Eqs. \eqref{xxix}, \eqref{xxx} and \eqref{xxxi} with $C_-=0$, $C_+=N$. 
It is convenient to solve equation \eqref{xxx} in the simplest gauge choice  $\rho=\phi$. This gauge choice uniquely fixes  the null coordinates $x^{\pm}$, in close analogy with the classical theory. These are commonly referred to as ``Kruskal coordinates'', mimicking the classical language. 

First we will briefly describe  the static, radiationless and asymptotically flat solutions. 
As explained in the previous section, the chiral functions $t_{\pm}$, in Kruskal gauge, are  fixed as 
\be 
t_{\pm}(x^{\pm})= \frac{1}{(2 x^{\pm})^{2}} \ . 
\ee
The semiclassical solution can be obtained immediately 
\be 
 e^{-2\phi}+{Na\over12}\phi=-\lambda^2x^+x^--\frac{N}{48}\log\left(-\lambda^2
x^+x^-\right)+\frac{M}{\lambda}\>,\label{xl}
\ee
where $M$ is  the mass of the solution above a reference point $M_0$; we take $M_0=0$ for simplicity. This vacuum solution is modified if we introduce a collapsing shell of matter \eqref{nullshell}.  The static solution becomes now a dynamical one
\bea
e^{-2\phi}+{Na\over12}\phi
&=&-\lambda^2x^+x^--\frac{N}{48}\log
\left(-\lambda^2x^+x^-\right)\nonumber\\
&&-{m\over\lambda x^+_0}\left(x^+-x^+_0
\right)\theta\left(x^+-x^+_0\right)+\frac{M}{\lambda}
\> . \label{xliii}
\eea
For a diagrammatic picture of the evaporation process, see
Fig.~\ref{PenroseRST}. This figure corresponds to the model with
$a = 1/2$ and is supplemented by a flat spacetime region beyond the
endpoint of the evaporation; this extension will be justified later.

For $x^+>x^+_0$ we have an evaporating black hole solution with a curvature singularity, $R\to \infty$,  lying precisely on the critical line $\Omega^{\prime}=0$, where $\Omega\equiv e^{-2\phi} + \frac{aN}{12} \phi$.\footnote{In the gauge $\rho =\phi$, the curvature takes the form $R=-8e^{-2\rho}\frac{\Omega^{''}}{(\Omega')^3} \partial_+\Omega\partial_-\Omega  $.} The curvature singularity is situated at \footnote{The location of the curvature singularity can equivalently be characterized by $e^{-2\phi_c}=\frac{Na}{24}$, implying that $\Omega_c=\alpha$.}
\be
\alpha=-\lambda^2x^+\left(x^-+\Delta\right)-\frac{N}{48}\log\left(
-\lambda^2x^+x^-\right)+{(m+M)\over\lambda}\>,\label{xliv}
\ee
where $\Delta={m\over\lambda^3x_0^+}$ and $\alpha={Na\over24}\left(1-\log
{Na\over24}\right)$. The critical line initially lies behind an apparent
horizon $\partial_+\phi=0$,\footnote{In terms of $\Omega$, note that $\partial_+\Omega=\Omega'(\phi)\partial_+\phi$.} which is given by the curve
$-\lambda^2x^+\left(x^-+\Delta\right)=\frac{N}{48}\>$.
Due to Hawking emission the apparent horizon recedes and intersects the
curvature singularity in a finite proper time. The intersecting point $(x^{+}_{\text{int}},x^{-}_{\text{int}})$
is located at
\be
x^-_{\text{int}}={-\Delta\over 1-\frac{N}{48}\exp\left[-\frac{N}{48}
\left({(m+M)\over\lambda}-\alpha\right)-1\right]}\>,\label{xlvi}
\ee
\be
x^+_{\text{int}}={1\over\lambda^2\Delta}\left\{\exp\left[{48\over N}
\left({(m+M)\over\lambda}-\alpha\right)+1\right]-\frac{N}{48}\right\}
\> . \label{xlvii}
\ee
After the apparent horizon reaches the critical line, the continuation of the dynamical solution would lead to a naked singularity. A way to address this problem, following the construction proposed in \cite{RST}, is to observe that the evaporating solution, when restricted to $x^-=x^-_{\text int}$, remarkably coincides with a static vacuum configuration of the form 
\be
e^{-2\phi}+{Na\over12}\phi=-\lambda^2x^+\left(x^-+\Delta\right)
-\frac{N}{48}\log\left(-\lambda^2x^+\left(x^-+\Delta\right)\right)+\frac{\hat M}{\lambda}
\>.\label{xlviii}
\ee
where the constant $\hat M$ is fixed as
\be
\frac{\hat M}{\lambda}=-\frac{N}{48}\left(1-\log\frac{N}{48}\right)+\alpha\> \ .\label{xlix}
\ee
Therefore, one can, albeit in a somewhat ad hoc manner, eliminate the region beyond $x^-=x^-_{\text int}$ in the original evaporating solution and replace it by gluing the above static solution along $x^-=x^-_{\text int}$. In any case, the persistence of the curvature singularity in the semiclassical regime will imply non-unitarity of the evaporation process. We now turn to the energy balance.

\subsection{Outgoing Hawking flux}

The Hawking flux is obtained by following the same strategy as in the classical case. We first determine the asymptotic form of the semiclassical solution and of the flux $\langle T^{f}_{x^-x^-}\rangle$ as $x^+\to \infty$ and then rewrite the result in terms of the affine coordinate $\hat \sigma^-$.
In the asymptotic region $x^+\to +\infty$, but now with $-\infty< x^-<x^-_{\text{int}}<x^-_h$, we can approximate the solution as 
\be \label{asymtoticsemiclassicalsolution}
e^{-2\phi} \approx -\lambda^2x^+(x^--x^-_h)  . \ee
The leading asymptotic behavior coincides with that of the classical solution.
Therefore, the calculation of $\langle T^f_{x^-x^-}\rangle$, for $x^+\to +\infty$, yields  the same analytical expression as in \eqref{fluxx-}  
\be 
\langle T^f_{x^-x^-}\rangle = -{N\over12}\left[\left(\partial_{-} \rho\right)^2-\partial^2_{-} \rho + t_{-}\right] \sim \frac{N}{48}\left[\frac{1}{(x^- - x^-_{h})^2} - \frac{1}{(x^-)^2}\right]
\ . \ee
The Hawking flux  measured by an inertial frame at $x^+\to +\infty$
is given by 
\be 
\langle T^f_{\hat \sigma^- \hat\sigma^-}\rangle= \left (\frac{dx^-}{d \hat\sigma^-}\right)^2 \langle T^f_{x^-x^-}\rangle \ , 
\ee
where $\frac{ dx^-}{d\hat\sigma^-}$ at $x^+\to +\infty$ can also be determined by solving the  geodesic equation.
Working in the asymptotic region $x^+\to +\infty$, one gets (recall the arguments of subsection \ref{outgoingflux})
\begin{equation}
    \frac{dx^-}{d\hat \sigma^-}= -\lambda(x^--x^-_h)
\ . \end{equation}
We find 
\bea \label{scriIfluxRST}\langle T^f_{\hat \sigma^-\hat\sigma^-}\rangle
&=&\lambda^2(x^--x^-_h)^2 \frac{N}{48}\left[\frac{1}{(x^- - x^-_{h})^2} - \frac{1}{(x^-)^2}\right] \nonumber \\
&=&\frac{N \lambda^2}{48}\left[1-(1+\frac{P}{\lambda} e^{\lambda \hat \sigma^{-}} )^{-2}\right] \ , \eea
with $P=m/(\lambda x^+_0)$ for the shock wave profile.

Note that the analytic expression for the Hawking flux, including backreaction effects, coincides with that obtained in the fixed-background approximation.\footnote{If we evaluate the flux for finite $x^+_r$ one finds a mismatch between the classical and the semiclassical calculation. In the limit $x^+_r \to +\infty$, the difference disappears.} The crucial difference is that, in the present case, we  cannot reach $ \hat \sigma^- \to +\infty$, since the evolution encounters the curvature singularity at $\hat \sigma^-_{\text{int}}$, corresponding to $x^-_{\text{int}}$, given by \eqref{xlvi}. This has an important physical consequence. The total energy radiated by the evaporating black holes is finite, in contrast to the result obtained in the absence of backreaction, and is given by
\bea \label{Erad}
E_{\text{rad}}&&=\int^{\hat\sigma^-_{\text{int}}}_{-\infty}\langle T^f_{\hat \sigma^- \hat \sigma^-}\left(\hat\sigma^-
\right)\rangle\,  d\hat\sigma^-\nonumber\\&&=m+M-\lambda\alpha
-{N\lambda\over48}\left(\log\frac{N}{48}-1\right)-{N\lambda\Delta\over48
x^-_{\text{int}}}\> \ . 
\eea

At this point, one may ask whether the evolution is consistent with energy conservation. To address this question properly, one must also take into account the presence of a thunderpop emitted from the endpoint of the evaporation at $\hat \sigma_{\text{int}}^-$. Indeed, the semiclassical constraint equation for the $(--)$ component [see eq. \eqref{xxxi}] implies the existence of a shock wave originating at the endpoint, which is given by
\be
T^f_{--}\left(\hat\sigma^-\right)={N\lambda\Delta\over48x^-_{\text{int}}}\,
\delta\left(\hat\sigma^--\hat\sigma^-_{\text{int}}\right)\> . \label{l}
\ee
The energy of the thunderpop can be obtained by integration and is straightforwardly given by 
\be E_{thp}= {N\lambda\Delta\over48x^-_{\text{int}}} \ . \ee

Therefore, the total energy of the initial configuration, $M$, together with the energy $m$ of the collapsing shell, must equal the sum of the radiated energy $E_{\text{rad}}$, the energy carried by the thunderpop, 
and the energy of the remnant, $\hat M$.
One can readily verify, using the result of the integral in Eq.~\eqref{Erad}, that energy conservation is exactly satisfied \footnote{A different prescription, in which the local counterterms are included in the radiated flux, would spoil this energy balance. Further support for the prescription adopted here can be found in Ref. \cite{Alexandre:2025}.}
\be M + m =  E_{\text{rad}} +  {N\lambda\Delta\over48x^-_{\text{int}}} +  \hat M \ . \ee

\begin{figure}
\begin{center}
\includegraphics[angle=0, width=80mm]{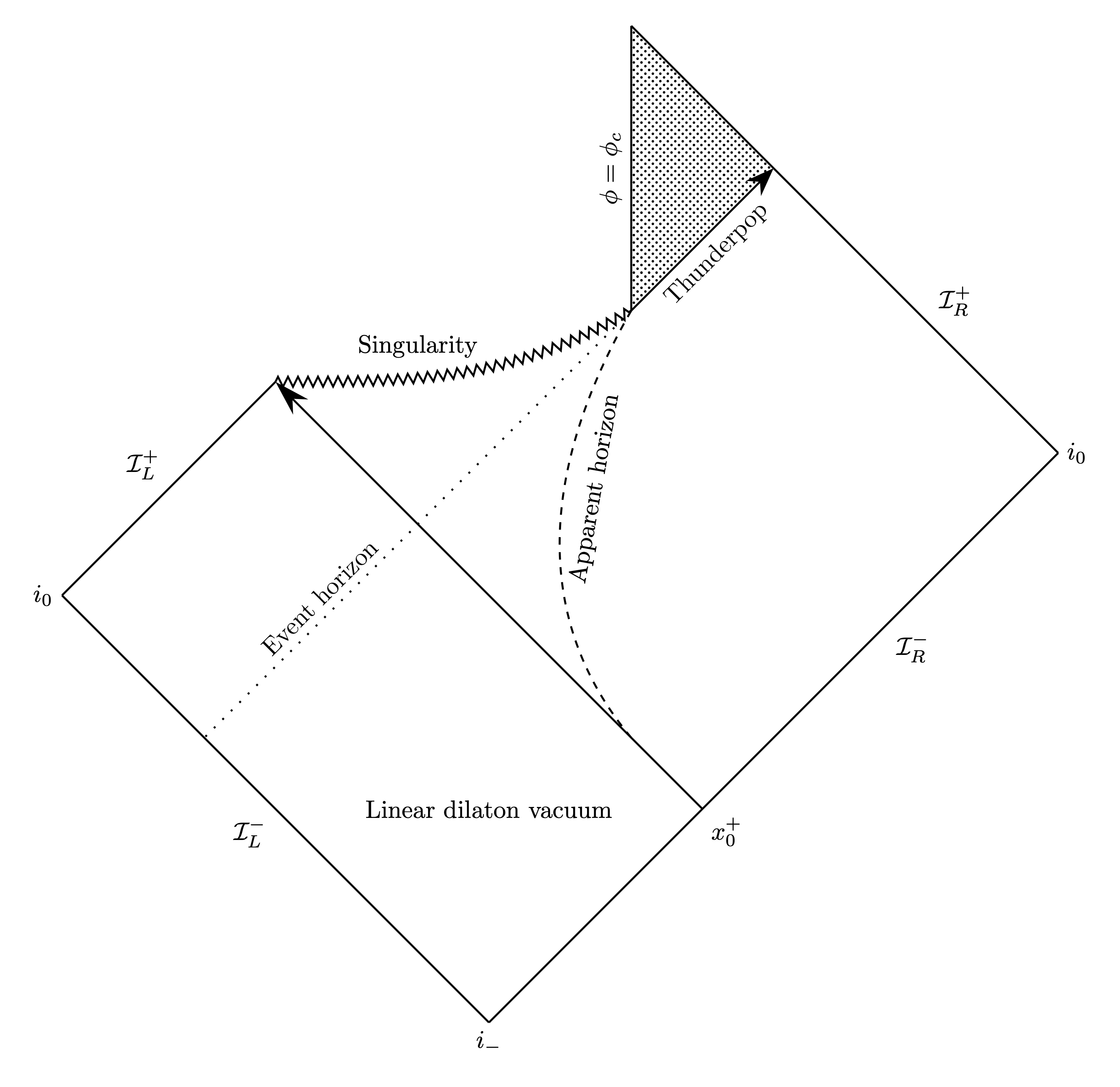}\end{center}
\caption{\small  Penrose diagram of an evaporating black hole in the semiclassical RST model.}
\label{PenroseRST}
\end{figure}

We note that the models with $a=1/2$ and  $a=0$ are particularly preferred, since in both cases the initial and final configurations coincide,  $M=\hat M$.  Moreover, for  $a=1/2$ the ground state is Minkowski space (the so-called linear dilaton vacuum), whereas for  $a=0$ it is a geodesically complete semi-infinite throat \cite{BPP}. The overall evaporation process for  $a=1/2$ is depicted in Fig.~\ref{PenroseRST}.

However, this does not mark the end of the story  since this formal energy balance does not capture all the physical features of the solution. In particular,  the matching conditions imposed at the null line $x^-= x^-_{\text{int}}$ are further supplemented by restricting the spacetime  at $x^- \geq x^-_{\text int}$ to the region to the right of the critical, timelike curve $\phi= \phi_{c}$, which constitutes the analytic continuation of the apparent horizon \cite{RST}.

The discontinuity in spacetime, illustrated in Fig. \ref{PenroseRST} through the Penrose diagram for the RST model, inevitably leads to an infinite burst of energy at the endpoint of evaporation, in addition to the finite energy `thunderpop" obtained before. This is indeed a general result, independent of any specific model. The key factor is the presence of the spacetime discontinuity caused by the singularity, which is ultimately responsible for the thunderbolt. A clear and accessible explanation of this effect can be found in \cite{FN05}. The phenomenon was first identified in a different physical context in Ref. \cite{AdW86}, which showed that, even in flat two-dimensional spacetime, a ``trousers"-like topology, such as the one shown in Fig. \ref{PenroseRST}, leads to infinite particle and energy production,  thereby violating energy conservation. As a result, the following three features appear to come together as a unified package:
i) the emergence of a spacelike curvature singularity,
ii) the breakdown of unitary evolution, and
iii) the violation of energy conservation.

The considerations discussed above justify the approach adopted in this paper. Avoiding the curvature singularity by any reasonable mechanism may allow for the preservation of unitary evolution, and perhaps even point towards the conservation of energy,  which however may prove to be a considerably more elusive issue. One of the key lessons from the analysis of the semiclassical equations of two-dimensional dilaton gravity is the emergence of subcritical regimes, in which no black hole forms and unitarity is maintained~\cite{ST94, DM94, CV94}. In certain situations, as one approaches the critical threshold for black-hole formation (for instance, when the mass of the incoming matter falls below a critical value), the emitted radiation may exhibit properties that, in some regions, are indistinguishable from those of Hawking thermal radiation~\cite{BPP96}. These insights motivate the exploration of a different type of ``subcritical'' regime. In the present work, this ``subcritical'' regime is defined by a negative total central charge of the system, arising from the combined contributions of the matter and ghost-gravity sectors.

\section{Backreacted solutions in the negative central charge regime}
\label{sec:backreacted solutions negative C}

Let us analyze in more detail the equations implied by our one-loop theory   \eqref{CGHSoneloop}.
To this end it is useful to reexpress $\Gamma_-[\hat g]$ in terms of $g$ and $\phi$. Since $\hat g_{\mu\nu}=e^{-2\phi}g_{\mu\nu}$ is a conformal transformation, we have
\bea  
-\frac{C_-}{96\pi}\int d^2x \sqrt{-\hat g}R(\hat g)\Box^{-1}(\hat g) R(\hat g) &=&  -\frac{C_-}{96\pi}\int d^2x \sqrt{-g}R(g)\Box^{-1}(g) R(g) \nonumber \\
&+& \frac{C_-}{24\pi}\int d^2 x \sqrt{-g}[(\nabla \phi)^2 -\phi R]\ . 
\eea
Therefore, the one-loop action \eqref{CGHSoneloop} becomes
\bea \label{oneloop2}  S=\, S_{\text{CGHS}}[g, \phi, f]  &-&\frac{(C_++C_-)}{96\pi}\int d^2x \sqrt{-g}R(g)\Box^{-1}(g) R(g)\nonumber \\
&+& \frac{C_+}{24\pi} \int d^2 x \sqrt{-g} [(1-2a) (\nabla \phi)^2 + (a-1)\phi R]  \nonumber \\
&+& \frac{C_-}{24\pi}\int d^2 x \sqrt{-g}[(\nabla \phi)^2 -\phi R]\ . \eea

The corresponding one-loop semiclassical equations are given by Eqs. \eqref{xxix}, \eqref{xxx} and \eqref{xxxi}. One of them is the free field equation $\partial_+\partial_-(\rho -\phi)=0$, since we have designed the overall action to fulfill it.  In Kruskal gauge $\rho = \phi$, and for the boundary conditions $t_{\pm}(x^{\pm}) = 1/(2x^{\pm})^2= \hat t_{\pm}$ the static solutions take the form 
\be
 e^{-2\phi}+{C_+a\over12}\phi=-\lambda^2x^+x^--{(C_++C_-)\over48}\log\left(-\lambda^2
x^+x^-\right)+\frac{M}{\lambda}\> . \label{xl2}
\ee
We require two-dimensional Minkowski spacetime to remain an exact one-loop solution. This condition uniquely fixes the parameter $a$ to $ a=a_0=(C_++C_-)/2C_+$,  as anticipated in \eqref{a=}. With this choice, the Minkowski background is recovered in the expected limit $M=0$, for which the one-loop solution simplifies to the compact expression
\be
 e^{-2\phi}+\frac{C_++C_-}{24}\phi
 =-\lambda^2x^+x^-
 -\frac{C_++C_-}{48}\log\!\left(-\lambda^2x^+x^-\right)\, .
 \label{xl22}
\ee
It is straightforward to verify that this equation admits the elementary solution $e^{-2\phi}=-\lambda^2 x^+x^-$, thereby ensuring that the linear dilaton vacuum remains an exact one-loop solution.

At the level of the action the choice $a=a_0$ reduces theory \eqref{CGHSoneloop} to
\bea 
\label{oneloop4} 
S= S_{\text{CGHS}}[g, \phi, f]
&-&\frac{(C_++C_-)}{96\pi}\int d^2x \sqrt{-g}\,R(g)\,\Box^{-1}(g)\,R(g)\nonumber \\
&-& \frac{C_+ + C_-}{48\pi}\int d^2 x \sqrt{-g}\,\phi R\ ,
\eea
which turns out to be the RST action with $N \to C_-+C_+$.
Therefore, although the physical reasoning followed here differs, we arrive at one-loop semiclassical equations that are mathematically equivalent to those studied in Refs.~\cite{PSS22,
PSS23, PSS23b}.

\subsection{Regime $C_-+C_+<0$}

We now consider the regime with $C_-+C_+ < 0$. We assume a given number $N = C_+$ of matter fields, with $N$ below the absolute value of the negative central charge $|C_-|$.\footnote{Note that one can also consider a model with $C_+ = N\hat N$ (fields $f_i$, $i = 1, 2, \dots, N\hat N$) that includes an additional conformal matter sector with negative central charge $c = -24\hat N$ coupled to the metric $\hat g$. Then $C_- = -24\hat N$. The large-$N$ limit is replaced here by the large-$\hat N$ limit, thus justifying the conventional semiclassical approach for a theory with total negative central charge. This mechanism (without the local counterterm $S_{local}(g,\phi)$) was also proposed in \cite{Strominger92}. For simplicity, we take $\hat N=1$.} We also consider a dynamical scenario with incoming classical matter given by a shock wave $T_{++}^{f} = \frac{m}{\lambda x_0^+}\, \delta(x^+ - x_0^+)$.

The vacuum both before and after \(x_0^{+}\) is described by the state \(\lvert 0_R \rangle\) for the ghost--gravity sector, corresponding to the choice \(\hat t_{\pm} = 1/(2x^{\pm})^{2}\). For the matter fields \(f_i\), the vacuum is assumed to coincide with the standard Minkowski vacuum \(\lvert 0_M \rangle\) in the region \(x^{+} < x_0^{+}\). After the passage of the shock wave, the vacuum is further assumed to remain Minkowskian at past null infinity, \(x^{-} \to -\infty\), which again implies \(t_{\pm} = 1/(2x^{\pm})^{2}\).

Using the shock-wave profile for the collapsing matter, we obtain the one-loop semiclassical solution
\be
 e^{-2\phi} + \frac{C_++C_-}{24}\,\phi
 = -\lambda^2 x^+ x^- - \frac{C_++C_-}{48}\,
   \log\!\left(-\lambda^2 x^+ x^-\right)
 - \frac{m}{\lambda x_0^+}\,(x^+ - x_0^+)\,\theta(x^+ - x_0^+)\ .
 \label{xl2a}
\ee
Introducing 
\be \kappa \equiv \frac{(C_+ + C_-)}{24}=\frac{N-24}{24} < 0 \ ,
\ee 
and $P \equiv m/(\lambda x_0^+)$, the expression above can be rewritten, after the shock wave, as
\begin{equation}
\label{Omegaeq}
    \Omega \equiv e^{-2\phi} + \kappa\,\phi
    = -\lambda^2 x^+\!\left(x^- + P/\lambda^2\right) + \frac{m}{\lambda}
      - \frac{\kappa}{2}\,\log\!\left(-\lambda^2 x^+ x^-\right)\ .
\end{equation}
Before the shock wave (i.e., for $m = P = 0$), the solution describes Minkowski spacetime. With respect to the metric $\hat g$, it instead corresponds to a right Rindler wedge. After the shock wave, since $d\Omega/d\phi \neq 0$, the apparent-horizon condition $\partial_+\phi=0$ translates into $\partial_+\Omega=0$. This gives 
\be x^-_{\rm ah}(x^+) = x_h^- -\frac{\kappa}{2\lambda^2 x^+}, \qquad x_h^- \equiv -\frac{P}{\lambda^2}. \ee 
Since $\kappa<0$, the apparent horizon lies to the right of the null line $x^-=x_h^-$ and approaches it asymptotically as $x^+\to+\infty$. The limiting null surface $x^-=x_h^-$ therefore plays the role of the event horizon. 

Evaluating the apparent-horizon curve at the shock wave gives \footnote{Our Kruskal chart is restricted to $x^+ > 0$, $x^- < 0$, since the logarithm in $\Omega$ is defined only there. Naively, a positive value of $x_{h_0}^-$ seems hard to interpret; for now we admit this possibility, and its physical justification will be given below.
}
\be x^-_{h_0} \equiv x^-_{\rm ah}(x_0^+) = x_h^- -\frac{\kappa}{2\lambda^2 x_0^+} = -\frac{1}{\lambda^2 x_0^+} \left( \frac{\kappa}{2} + \frac{m}{\lambda} \right). \ee
In contrast to the regime analyzed in the previous section, the trapped region does not disappear. Moreover, the null surface $x^-=x_h^-$ has the same infinite-redshift character as in the classical theory: as $x^-\to x_h^-$, a finite interval $\Delta\sigma^-$ is mapped into a divergent interval $\Delta\hat\sigma^-$.

\begin{figure}
\begin{center}\includegraphics[angle=0, width=65mm]{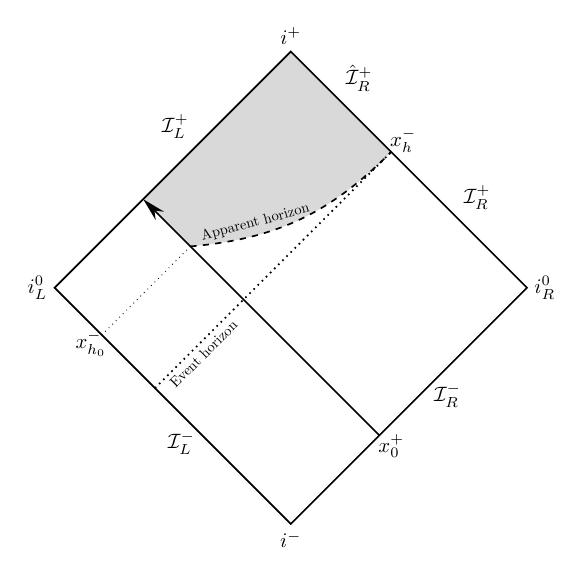}\end{center}
\caption{\small Penrose diagram for gravitational collapse induced by a shock wave.
In this model, an apparent horizon forms (dashed line) but does not 
contract under Hawking radiation. The straight dashed lines correspond to null geodesics at finite coordinate $x^\pm$.}
\label{PenroseNegativeC}
\end{figure}

One can easily check that the scalar curvature $R = 8 e^{-2\rho}\,\partial_+ \partial_- \rho$ is nowhere singular and that it vanishes along any null direction. In the exterior region, the quantity $r \equiv e^{-\phi}$ diverges as $x^+ \to +\infty$.\footnote{$e^{-\phi}$ can be interpreted as a radial-type coordinate, following the standard analogy in which $e^{-2\phi}$ represents the area of the two-spheres.} Inside the horizon (i.e., for $x^+<x^-_h<0$), the ``radial'' coordinate $r \equiv e^{-\phi}$ tends to zero as $x^- \to 0$ at fixed $x^+$ and as $x^+ \to +\infty$ at fixed $x^-$. In this regime, however, both asymptotic regions near $r = 0$ are flat. The overall picture is represented in the Penrose diagram of Fig.~\ref{PenroseNegativeC}.

The above discussion agrees with the spacetime picture already described in \cite{PSS23b}. The main differences with respect to that approach will become apparent in the following subsections, where we study the radiation fluxes and their correlations in the asymptotic regions.

\subsection{Fluxes}

\begin{figure}  \begin{center}\includegraphics[angle=0, width=80mm]{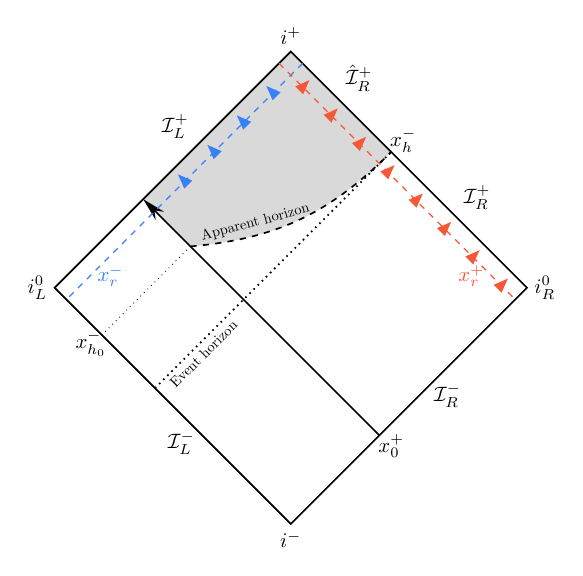}\end{center}\caption{Ingoing and outgoing energy fluxes}\label{fig:Hawkingflux}\end{figure}

In this subsection we discuss how to compute the ingoing and outgoing fluxes induced by the collapsing matter in the semiclassical geometry obtained above. We focus here on the numerical analysis based on the exact solution, which can be expressed in terms of the Lambert function $W(\Omega)$:
\be
\label{exactsolution1}
\rho = \phi = \frac{\Omega}{\kappa} + \frac{1}{2}\,W,
\qquad
W \equiv W_0\!\left(-\frac{2}{\kappa}\,e^{-2\Omega/\kappa}\right),
\ee
where
\be
\label{exactsolution2}
\Omega = -\lambda^2 x^+ (x^- - x_h^-) + \frac{m}{\lambda}
- \frac{\kappa}{2}\,\log(-\lambda^2 x^+ x^-)\,.
\ee
The corresponding exact Polyakov fluxes [see Eq.~\eqref{mastereq}]
can be written as
\be
\langle T^f_{x^\pm x^\pm}\rangle = -\frac{N}{12}\left[
\frac{1-W}{\kappa^{2}(1+W)^{3}}\,(\partial_\pm\Omega)^{2}
- \frac{\partial_\pm^{2}\Omega}{\kappa(1+W)}
+ \frac{1}{4(x^\pm)^{2}}\right]\,.
\ee
The physical fluxes are then obtained by introducing affine parameters $\hat\sigma^\pm$ along the corresponding null rays
$x^+ = x^+_r$ or $x^- = x^-_r$ (see Fig. \ref{fig:Hawkingflux})
\bea
\langle T^f_{\hat\sigma^- \hat\sigma^-}\rangle(x^-,x^+_r) &=&
\left(\frac{dx^-}{d\hat\sigma^-}\right)^{2}\!\Bigg|_{x^+_r}
\langle T^f_{x^- x^-}\rangle(x^-,x^+_r)\,, \label{eq:Hawkingflux}\\
\langle T^f_{\hat\sigma^+ \hat\sigma^+}\rangle(x^-_r,x^+) &=&
\left(\frac{dx^+}{d\hat\sigma^+}\right)^{2}\!\Bigg|_{x^-_r}
\langle T^f_{x^+ x^+}\rangle(x^-_r,x^+)\,.
\eea
The affine parameters $\hat\sigma^\pm$ are normalized by matching them to the inertial null coordinates $\sigma^\pm$ of the two-dimensional Minkowski region before the shock wave. For example, from the geodesic equation
\begin{equation}
\frac{d^{2}x^+}{d\hat\sigma^{+\,2}}
+ \Gamma_{++}^{+}\,\frac{dx^+}{d\hat\sigma^+}\,
                   \frac{dx^+}{d\hat\sigma^+} = 0\,,
\end{equation}
and using $\Gamma_{++}^{+} = 2\,\partial_+\rho$, we obtain the
general solution along the null geodesic $x^- = x^-_r$,
\begin{equation}
\label{sigmax}
\frac{dx^+}{d\hat\sigma^+} = e^{-2\rho(x^+, x^-_r) + B(x^-_r)}\,.
\end{equation}
Here, the functional dependence of the integration constant $B(x^-_r)$ is determined by requiring that, at $x^+ = x^+_0$, the affine parameter $\hat\sigma^+$ joins smoothly onto the corresponding coordinate $\sigma^+$ defined prior to the shock wave.\footnote{As a direct consequence, the energy of the shock wave measured at $\mathcal{I}^+_L$ equals $m$, in agreement with its value in our reference frame at past null infinity.} We then have the relation
\be
\frac{dx^+}{d\hat\sigma^+}\Bigg|_{x^+=x^+_0}
= \frac{dx^+}{d\sigma^+}\Bigg|_{x^+=x^+_0}
= \lambda x^+_0\,.
\ee
From~\eqref{sigmax}, on the other hand,
\begin{equation}
\frac{dx^+}{d\hat\sigma^+}\Bigg|_{x^+=x^+_0}
= -\lambda^{2} x^+_0\, x^-_r\, e^{B(x^-_r)}\, ,
\end{equation}
as a consequence of continuity of the metric $\rho$ at the shockwave. This yields $B(x^-_r) = -\log(-\lambda x_r^-)$, so that
\begin{equation}\label{x+hatsigma+}
\frac{dx^+}{d\hat\sigma^+} = \frac{1}{-\lambda x^-_r}\,
e^{-2\rho(x^+, x^-_r)}\,.
\end{equation}
A similar calculation for the other chiral sector yields
\be
\frac{dx^-}{d\hat\sigma^-}\Bigg|_{x^+_r}
= \frac{1}{\lambda x_r^+}\,e^{-2\rho(x_r^+, x^-)},
\qquad
\frac{dx^+}{d\hat\sigma^+}\Bigg|_{x^-_r}
= \frac{1}{-\lambda x_r^-}\,e^{-2\rho(x^+, x_r^-)}\,.
\ee
\\

\begin{figure} 
\includegraphics[angle=0, width=99mm]{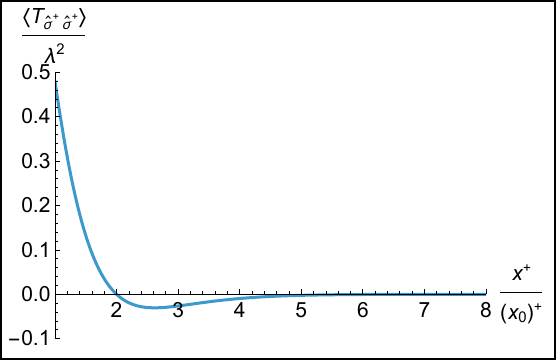}
\caption{Flux at the left null infinity $\mathcal{I}_L^+$ for $x^+>x^+_0$, computed from the parameter set $N=23$, $m/\lambda=1/48$, $\lambda x^-_h=-1$ in units with $\hbar=1$; and therefore 
$\kappa = -1/24$.}\label{fig:L+}
\end{figure}

We now have all the exact expressions needed to perform the numerical computations. A typical ingoing flux reaching $\mathcal{I}_L^+$ is shown in Fig.~\ref{fig:L+}, where, for convenience, we have set the parameters to  
$N=23$, $m/\lambda=1/48$, $\lambda x^-_h=-1$ in units with $\hbar=1$. These values place us in the central-charge regime with $\kappa = -1/24$. The horizontal axis is the dimensionless ratio $x\equiv x^{+}/x_{0}^{+}$. The plot shows a positive flux, that shortly after turns negative and reaches a minimum, and later becomes exponentially suppressed.
\

\begin{figure}[!h]  

\includegraphics[scale=0.9]{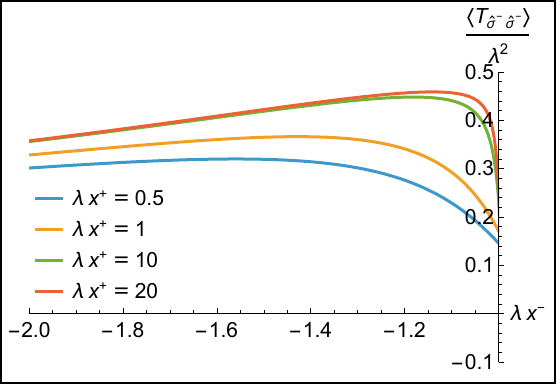}
\includegraphics[scale=0.91]{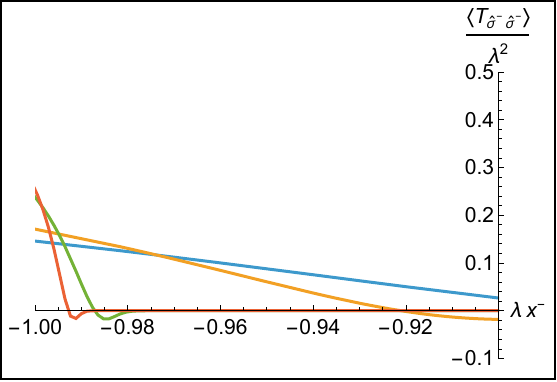}

\caption{Flux at increasingly high values of $\lambda x^+\equiv \lambda x^+_r$. The plots represent the exterior ($x^-<x^-_h$) and interor ($x^->x^-_h$) regions respectively, and are computed for the parameter set
$N=23$, $m/\lambda=1/48$, $\lambda x^-_h=-1$ in units with $\hbar=1$; and therefore
$\kappa = -1/24$.}
\label{R+}
\end{figure}

The corresponding outgoing flux at the surface $x^+ = x^+_r$ is shown in Fig.~\ref{R+} for increasing values of $x^+_r$.
Three qualitative facts deserve emphasis. First, the flux is positive everywhere in the exterior region, i.e., the asymptotic observer at $\mathcal I_R^+$ sees a genuine outgoing energy flux at every $x^{-}$ to the left of the horizon $x^-=x^-_h$, for every $x^-$ in the explored range; the negative dip immediately to the right of $x^-_h$  is exponentially
suppressed. 
Second, the flux saturates as $x^+_r\to+\infty$: the curves for $\lambda x^+=10$ and $\lambda x^+=20$ are visually indistinguishable away from the horizon.  Third, all curves develop a peak slightly to the left of the horizon, with the peak height growing weakly with $x^+_r$.

To complement our analysis we will consider the relevant  asymptotic regimes, which naturally follow  from the standard expansions of the Lambert function:
\be
W_0(z)=z+\mathcal O(z^2)\qquad (z\to0),
\ee
and
\be
W_0(z)=\log z+\mathcal O\!\left(\log\log z\right)
\qquad (z\to+\infty).
\ee
Thus, when $\Omega\to-\infty$, one has $z\to0$ and therefore $W_0(z)\sim z$, which implies $\Omega\sim\kappa\rho$. In contrast, when $\Omega\to+\infty$, one has $z\to+\infty$, so that $W_0(z)\sim\log z$, which implies $e^{-2\rho}\sim\Omega$.  With these expansions we can compute analytically the behaviour of the fluxes in the relevant asymptotic regions:

\begin{enumerate}
  \renewcommand{\labelenumi}{\roman{enumi}.}

\item {\bf Ingoing flux on $\mathcal I_L^+$.}  For $x^+ < x^+_0$ (before the shock wave) there is no radiation. The novelty of our model is that it now makes sense to ask about the flux through $\mathcal{I}^+_L$ for $x^+ > x^+_0$, and the result is non-trivial. Taking $x_r^-\to0^-$, and with $x^+>x_0^+$ fixed, one has $\Omega\to-\infty$, hence $\rho\sim \kappa^{-1}\Omega$.
  In this limit, the affine factor along the null ray $x^-=x^-_r$ becomes
\be 
\left.\frac{dx^+}{d\hat\sigma^+}\right|_{\mathcal I_L^+}
=
\lambda x^+
\exp\!\left[
-\frac{2m}{\kappa\lambda}
\left(1-\frac{x^+}{x_0^+}\right)
\right] \ , \ee
and the ingoing flux is 
\be \label{eq:ingoingfluxLp}
\langle T^f_{\hat\sigma^+ \hat\sigma^+}\rangle_{\mathcal I_L^+}
= -\frac{N \lambda m}{12\,\kappa^{2}}\,\frac{x^+}{x^+_0}
   \!\left(\kappa + \frac{m}{\lambda}\,\frac{x^+}{x^+_0}\right)
   \exp\!\left[-\frac{4m}{\kappa\lambda}
   \!\left(1 - \frac{x^+}{x^+_0}\right)\right]\, ,
\ee
where we have used $x_h^{-} =-\frac{m}{\lambda^3 x_0^{+}}$. Note that the sign of this flux  is fixed by the factor $\bigl(\kappa + \tfrac{m}{\lambda}\,\tfrac{x^+}{x^+_0}\bigr)$. At $x^+ = x^+_0$ the flux is generically nonzero, and it decays exponentially as $x^+ \to +\infty$, owing to $\kappa < 0$. The resulting flux \eqref{eq:ingoingfluxLp} is shown in Fig.~\ref{fig:L+}.
  
    \item {\bf Outgoing flux on $\mathcal I_R^+$.} Taking $x_r^+\to+\infty$ with $x^-<x_h^-$ fixed, one has $\Omega\to+\infty$, hence $e^{-2\rho}\sim\Omega\sim \lambda^2 x^+(x^-_h-x^-)$. Using $x^{-}-x_h^{-}=-\lambda^{-1} e^{-\lambda \hat{\sigma}^{-}}$ and $x_h^{-}=-\lambda^{-2}P$, the exact expression in this limit reduces to 
\bea \label{eq:outgoingfluxR}
\langle T^f_{\hat\sigma^- \hat\sigma^-}\rangle_{\mathcal{I}^+_R}
&=& \frac{N\lambda^2}{48}(x^--x^-_h)^2\left[\frac{1}{(x^- - x^-_{h})^2} - \frac{1}{(x^-)^2}\right] \nonumber \\&=&\frac{N\,\lambda^{2}}{48}
  \left[1 - \!\left(1 + \frac{P}{\lambda}\,
   e^{\lambda \hat\sigma^{-}}\right)^{\!-2}\right] \ .
\eea
It is worth noticing that we obtain the same analytical expression for the Hawking flux of the $C_+ = N>0$ matter fields. The main difference from the RST-type models of the conventional semiclassical approach is that the null coordinate $\hat\sigma^-$ now ranges from $-\infty$ to $+\infty$, as in the classical collapse scenario discussed in Section~\ref{section2}. In this regime, the absence of a curvature singularity removes the intersection point $(x^+_{\mathrm{int}}, x^-_{\mathrm{int}})$ between the singularity and the apparent horizon that is characteristic of the RST-type models. In other words, $\scri_R^+$ is complete, and the causal past of $\scri_R^+$ has the boundary $x^- = x^-_h$, as in the classical solution. For this reason we can still refer to it as the event horizon. This leads to a drastic modification of the interior of the horizon. In particular, one may ask about the flux entering the horizon and terminating at the additional asymptotically flat boundary at $r \equiv e^{-\phi} \to 0$.

\item {\bf Outgoing flux on $\hat{\mathcal I}_R^+$.} Taking $x_r^+$ large with $x_h^-<x^-<0$ fixed, one has again $\Omega\to-\infty$, hence $\rho\sim \kappa^{-1}\Omega$. The leading order reduces to
\bea \label{eq:outgoingfluxIRp}
\langle T^f_{\hat\sigma^- \hat\sigma^-}\rangle(x^-, x^+_r)
&=& -\frac{N\lambda^{2}}{12}\,e^{-\frac{4m}{\lambda\kappa}}
   \!\left(\frac{\lambda^{2}\,x^+_r\,x^-}{\kappa}\right)\!
   \left[\frac{\lambda^{2}\,x^+_r\,x^-}{\kappa} + 1\right]
   \exp\!\left[\frac{4\lambda^{2}}{\kappa}\,
   x^+_r(x^- - x^-_h)\right]\,.
\eea
Under the approximation used here, this flux is  negative everywhere. This only partially agrees with the numerical results based on the exact solution. This means that the above approximation cannot be trusted in the near-horizon region  (the transition region), where a different analytical approximation is required. 

\item  {\bf Outgoing flux in the near-horizon region.} Setting $x^-=x_h^-+\epsilon$, the exact flux is regular at $\epsilon=0$. If one subsequently takes $x_r^+\to+\infty$, then $\Omega(x_r^+,x_h^-)\to+\infty$, so that $W_0(z)\sim\log z$ and $e^{-2\rho}\sim\Omega$. One then finds
\begin{equation}
    \langle T^f_{\hat \sigma^- \hat \sigma^-} \rangle(x^-,x^+_r)=\mathcal{F}_0+\epsilon\,  \mathcal{F}_1 + \mathcal{O}(\epsilon^2)
\end{equation}
where 
\begin{equation}
\mathcal{F}_0 \sim \frac{\lambda^{2} N}{48}\,,
\qquad
\mathcal{F}_1  \sim \frac{N \lambda^{4}\,x^+_r}
   {6\,\kappa\,\log^{2}\!\bigl(2\,\lambda^{2}\,\kappa^{-1}\,x_r^+\,x^-_h\bigr)}\ .
\end{equation}
Since $\mathcal{F}_1 < 0$, for $\epsilon > 0$ the asymptotic flux
vanishes at
\begin{equation}
\epsilon_{\text{zero}} = -\frac{\mathcal{F}_0}{\mathcal{F}_1}\,.
\end{equation}
In the strict limit $x^+_r \to \infty$, this zero coincides with the
horizon, $\epsilon_{\text{zero}} \to 0^+$, in agreement with the
numerical plots in Fig.~\ref{R+}.

In summary, as $x^+_r \to +\infty$ the local flux profile collapses onto a step-like configuration: it takes the constant value $N\lambda^{2}/48$ on the exterior side $x^- < x^-_h$ and drops sharply to zero at the horizon. The reason is that the redshift factor $dx^-/d\hat\sigma^-$ that enters the physical flux vanishes in this limit. This extreme behavior is responsible for the suppression of the Hawking flux in the interior region for $x^+_r \to +\infty.$ It also implies that the affine parameter $\hat\sigma^-$ is not well defined in the interior region in this limit, a fact probably related to the completeness of $\scri_R^+$. Conversely, the peak observed at finite $x^+_r$ in the exterior region is a transient, whose location merges with the horizon in the same limit.

\end{enumerate}

\subsection{Correlations}

\begin{figure} 
\includegraphics[angle=0, width=80mm]{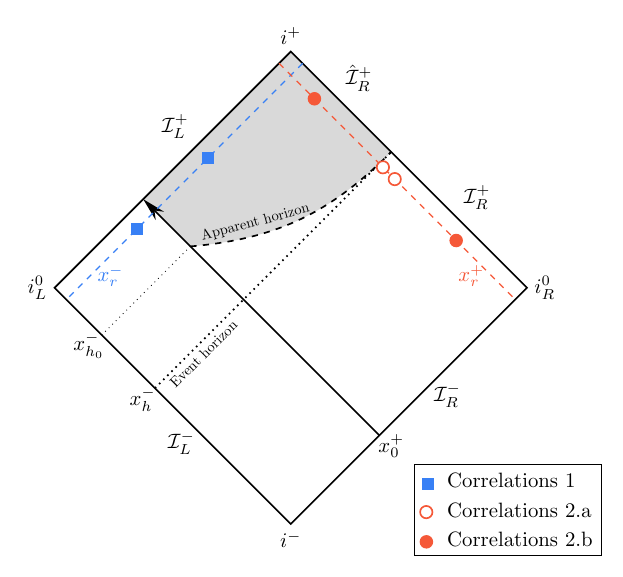}
\caption{
The figure shows representative pairs of points for which the corresponding energy-energy correlation functions
are analyzed in the text. We consider both the left- and right-moving sectors, with the corresponding pairs of points denoted by diamonds and circles, respectively.}\label{correlationsR}
\end{figure}

We now turn to the unitarity issue.  Owing to the absence of singularities in our spacetime, there is no geometrical obstruction to defining a unitary $S$-matrix from the asymptotically flat region $\mathcal{I}^-_R$ to the asymptotically flat region $\mathcal{I}^+_L$. To display this, and to analyze the behavior of the $S$-matrix from $\mathcal{I}^-_L$ to the surfaces $x^+ = x^+_r$ for increasing values of $x^+_r$, it is convenient to consider the energy-flux two-point functions
\be
 C_{\pm\pm,\,\pm\pm} = \langle T_{\pm\pm}(x)\,T_{\pm\pm}(x')\rangle
 - \langle T_{\pm\pm}(x)\rangle\,\langle T_{\pm\pm}(x')\rangle \ .
\ee
To evaluate these correlators, we use the identification introduced in Section~\ref{section2}:
\begin{equation} \label{TTno}
 T_{\pm\pm} = {:}T_{\pm\pm}{:} -\frac{N}{12}
 \left[(\partial_{\pm}\rho)^2 - \partial^2_{\pm}\rho\right]\,.
\end{equation}
Since we are interested in connected correlators, the contributions from the second term in~\eqref{TTno} cancel, leaving only correlators of the normal-ordered operators ${:}T_{\pm\pm}{:}$, which can be computed using Wick's theorem in the standard way for two-dimensional conformal field theory~\cite{CFTbook}. Therefore, the transformation laws of the connected correlators are given by
\begin{equation}\label{Ctransformation}
C_{\hat \sigma^\pm \hat \sigma^\pm, \ \hat \sigma^{\prime \pm} \hat \sigma^{\prime \pm}}(\hat \sigma^{\pm}, \hat \sigma^{\prime\pm})
=
\left(\frac{dx^\pm}{d\hat \sigma^\pm}\right)^2
\left(\frac{dx^{\prime \pm}}{d\hat{\sigma}^{\prime \pm}}\right)^2
C_{x^\pm x^\pm, \ x^\pm x^\pm}(x^{\pm}, x^{\prime \pm})
\end{equation}
We will also analyze these correlators in the appropriate asymptotic regions below and extract their physical consequences. For a general picture of the expected correlations, see Fig. \ref{correlationsR}

\subsubsection{Correlations in the left moving sector}

We now analyze the correlation functions $C_{++,++}$. At $\mathcal{I}^-_L$, the energy--energy correlations are given by the standard expression \cite{CFTbook} 
\be
C^{\rm in}_{\sigma^+\sigma^+,\,\sigma^{\prime+}\sigma^{\prime+}}
= \frac{N}{8}\,\frac{1}{(\sigma^{+}-\sigma^{\prime+})^{4}}
= \frac{N}{8}\,\frac{\lambda^{4}}
   {[\log(\lambda x^{+}) - \log(\lambda x^{\prime+})]^{4}}\,.
\ee
Evolving them to $x^- \to 0$, we obtain
\bea
\label{Coutin+}
C^{\rm out}_{\hat\sigma^+\hat\sigma^+,\,
              \hat\sigma^{\prime+}\hat\sigma^{\prime+}}
&=& \frac{N}{8}\,\frac{\lambda^{4}}
   {[\log(\lambda x^{+}) - \log(\lambda x^{\prime+})]^{4}}
   \left(\frac{d\sigma^+}{d\hat\sigma^+}\right)^{2}
   \left(\frac{d\sigma^{\prime+}}{d\hat\sigma^{\prime+}}\right)^{2}
   \nonumber \\
&=& \frac{N}{8}\,\frac{\lambda^{4}}
   {[\log(\lambda x^{+}) - \log(\lambda x^{\prime+})]^{4}}\,
   \frac{1}{(\lambda x^+)^{2}\,(\lambda x^{\prime+})^{2}}
   \left(\frac{dx^+}{d\hat\sigma^+}\right)^{2}
   \left(\frac{dx^{\prime+}}{d\hat\sigma^{\prime+}}\right)^{2}\,.
\eea
For $x^+, x^{\prime+} < x^+_0$ the correlations are preserved identically. For $x^+ < x^+_0$ and $x^{\prime+} > x^+_0$, the derivative factors are
\begin{equation} 
\label{Jin2}
\left(\frac{dx^{\prime+}}{d\hat\sigma^{\prime+}}\right)^{2}
= (\lambda x^{\prime+})^{2}\,
  \exp\!\left[-\frac{4m}{\kappa\lambda}
  \!\left(1 - \frac{x^{\prime+}}{x^+_0}\right)\right]\,,
\end{equation}
\begin{equation} \label{Jout}
\left(\frac{dx^{+}}{d\hat\sigma^{+}}\right)^{2}
= (\lambda x^+)^{2}\,.
\end{equation}
The ratio of the correlations for $x^+ < x^+_0$ and $x^{\prime+} > x^+_0$ is therefore
\begin{equation}\label{eq:decoh}
\frac{C^{\rm out}(x^+,x^{\prime+})}{C^{\rm in}(x^+,x^{\prime+})}
= \exp\!\left[-\frac{4m}{|\kappa|\,\lambda}
  \!\left(\frac{x^{\prime+}}{x^+_0} - 1\right)\right]\,.
\end{equation}
We see that the correlations on $\mathcal{I}_L^{+}$ are exponentially suppressed relative to their in-vacuum value as $x^{\prime+} \to +\infty$. This carries an important lesson: the theory remains unitary at any finite $x^{\prime+}$, but cross-correlations across the shock wave become increasingly difficult to detect as $x^{\prime+} \to +\infty$.

\subsubsection{Correlations in the right-moving sector}

The behavior of the energy-energy correlation function in the right-moving sector follows from the analogous discussion about the outgoing fluxes.
For $x^+_r < x^+_0$ (i.e., before the shock wave), the correlation function $C_{--,--}$ is given by
\be
C_{\sigma^-\sigma^-,\,\sigma^{\prime-}\sigma^{\prime-}}
= \frac{N}{8}\,\frac{1}{(\sigma^{-} - \sigma^{\prime-})^{4}}
= \frac{N}{8}\,\frac{\lambda^{4}}
   {[\log(-\lambda x^{-}) - \log(-\lambda x^{\prime-})]^{4}}\,,
\ee
and becomes, for $x^+_r > x^+_0$,
\bea \label{Coutin}
C_{\hat\sigma^- \hat\sigma^-,\,\hat\sigma^{\prime-}\hat\sigma^{\prime-}}
&=& \frac{N}{8}\,\frac{\lambda^{4}}
   {[\log(-\lambda x^{-}) - \log(-\lambda x^{\prime-})]^{4}}
   \left(\frac{d\sigma^-}{d\hat\sigma^-}\right)^{2}
   \left(\frac{d\sigma^{\prime-}}{d\hat\sigma^{\prime-}}\right)^{2} \nonumber \\
&=& \frac{N}{8}\,\frac{\lambda^{4}}
   {[\log(-\lambda x^{-}) - \log(-\lambda x^{\prime-})]^{4}}\,
   \frac{1}{(\lambda x^-)^{2}\,(\lambda x^{\prime-})^{2}}
   \left(\frac{dx^-}{d\hat\sigma^-}\right)^{2}
   \left(\frac{dx^{\prime-}}{d\hat\sigma^{\prime-}}\right)^{2}\,.
\eea
When $x^-$ and $x^{\prime-}$ both lie in the exterior of the apparent black hole, sufficiently close to $x^-_h$, the limit $x^+_r \to \infty$ yields the thermal correlation function\footnote{These results are consistent with those of~\cite{Wilczek93}.}
\be
C_{\hat\sigma^- \hat\sigma^-,\,\hat\sigma^{\prime}\hat\sigma^{-\prime-}}
= \frac{N\lambda^{4}}{8}\,
  \frac{e^{2\lambda|\hat\sigma^- - \hat\sigma^{\prime-}|}}
       {(e^{\lambda|\hat\sigma^- - \hat\sigma^{\prime-}|} - 1)^{4}}\,.
\ee
If, instead, $x^-$ lies in the exterior region and $x^{\prime-}$ in the interior, non-thermal correlations arise, but they decay exponentially as $x^+_r \to +\infty$. This can be shown in several ways. For example, using the analytical approximation introduced earlier for the redshift factors:
\begin{equation} \label{Jin}
\left(\frac{dx^{\prime-}}{d\hat\sigma^{\prime-}}\right)^{2}
= (-\lambda x^{\prime-})^{2}\,
\exp\!\left[\frac{4\lambda^{2}}{\kappa}\,x^+_r\,
(x^{\prime-} - x^-_h) - \frac{4m}{\lambda\kappa}\right]\,,
\end{equation}
\begin{equation} \label{Jout2}
\left(\frac{dx^{-}}{d\hat\sigma^{-}}\right)^{2}
= \left[-\lambda(x^- - x^-_h)
  - \frac{\kappa}{2\lambda\,x^+_r}\,
  \log\!\left(-\lambda^{2}\,x^+_r\,x^-\right)\right]^{2}\,.
\end{equation}
As $x^+_r \to +\infty$, the exterior factor~\eqref{Jout2} approaches the standard expression required for thermal correlations in the exterior region,
\begin{equation} \label{Joutbis}
\left(\frac{dx^{-}}{d\hat\sigma^{-}}\right)^{2}
= \bigl[-\lambda(x^- - x^-_h)\bigr]^{2}\,,
\end{equation}
whereas the interior factor tends to zero. Consequently, the correlation function decays exponentially in this limit. This leads to an important implication, already foreshadowed in the analysis of the left-moving sector. The correlations between quanta emitted on the two sides of the horizon gradually vanish as one approaches null infinity. The theory remains unitary at any finite distance from the black hole, but cross correlations across the horizon become increasingly difficult to detect as $x^+_r \to +\infty$.

\subsection{Energy balance}

In this subsection, we examine the energy balance following the discussion of Section~\ref{sectionLargeN}, where energy conservation in RST-type models was analysed within the conventional semiclassical framework. The principal difference here is the appearance of additional asymptotically flat regions at future null infinity, $\mathcal{I}^+_L$ and $\hat{\mathcal{I}}^+_R$, together with their associated ingoing and outgoing fluxes. The energy-balance condition for the model with total negative central charge therefore takes the form: the energy of the collapsing shell at $\mathcal{I}^-_R$, equal to $m$, must equal the sum of the radiated energy at $\mathcal{I}^+_L$, $E^L_{\text{rad}}$, the energy carried by the shell as measured at $\mathcal{I}^+_L$ (also $m$), and the radiated energy in the right sector,
\begin{equation}
    m = E^L_{\text{rad}} + m + E^R_{\text{rad}}\,.
\end{equation}
The quantity $E^L_{\text{rad}}$ admits a closed-form analytic expression, 
\begin{equation}
  E_{\text{rad}}^{L}
  = \int_{x^+_0}^{\infty}\!\!\langle T^f_{x^+ x^+}\rangle\,
    \left (\frac{dx^+}{d\hat\sigma^+}\right)\,dx^+
  = \frac{N}{48}\!\left(\lambda + \frac{2m}{\kappa}\right)\,.
\end{equation}
Note that the sign of $E^L_{\text{rad}}$ is governed by the factor $\bigl(\lambda + \tfrac{2m}{\kappa}\bigr)$, which also determines whether the apparent horizon forms at the shock wave or only subsequently.\footnote{Specifically, if $\lambda + \tfrac{2m}{\kappa} < 0$, the apparent horizon emerges at the shock-wave location    ($x^+ = x^+_0,\quad x^- = x^-_h - \tfrac{\kappa}{2\lambda^{2} x^+_0}$), whereas if $\lambda + \tfrac{2m}{\kappa} \geq 0$, it forms later at the asymptotic point ($x^+ = -\tfrac{\kappa\lambda x^+_0}{2m},\quad x^- = 0$).} We further remark that $E^L_{\text{rad}}$ is obtained by integrating the stress-energy tensor along the null geodesic $x^- = 0$; since this integral can in general be negative, it signals a violation of the average null energy condition (see the review~\cite{Kontou2020}).

The evaluation of $E_{\text{rad}}^R$ is considerably more subtle. It is obtained by computing the total energy flux across the null surface $x^+ = x^+_r$ and then taking the limit $x^+_r \to +\infty$; only in this asymptotic regime, if at all, can energy conservation be recovered. This is arguably the most challenging aspect of our analysis. We recall that, within the conventional semiclassical framework of Section~\ref{sectionLargeN}, a subtle violation of energy conservation was found to be associated with the curvature singularity. In the present model, the absence of such a singularity in the regime of negative total central charge alleviates concerns regarding unitary evolution. Nevertheless, we are not yet able to establish energy conservation. The asymptotic behaviour of $E^R_{\text{rad}}$ is straightforward to obtain,
\begin{equation}
E^R_{\text{rad}} \sim \frac{N}{48}\,\lambda\,
\ln\!\frac{x_r^{+}}{x^+_0}\,,
\end{equation}
and this result is incompatible with the energy-balance condition proposed above. The origin of this unbounded growth can be traced to the persistence of the event horizon even after backreaction effects are incorporated: the total negative central charge proves sufficient to remove the curvature singularity, yet leaves the event horizon intact. Indeed, $\mathcal{I}^+_R$ is found to be complete, which is precisely why the event horizon persists.

It was shown in Ref.~\cite{dRMN25} that, for the spherically reduced model of Einstein gravity, a total negative central charge can remove both the singularity and the event horizon. The difference between the two situations lies in the independence of the two sectors. In the two-dimensional black hole the left and right sectors are entirely disconnected, whereas in four dimensions they are coupled through the timelike axis of symmetry $r = 0$, which acts as a mirror relating both chiral sectors via reflecting boundary conditions. This coupling between the two chiral sectors may be the missing key feature of the present model and the ultimate reason why a proper energy balance cannot be established, despite preserving unitarity in each chiral sector independently, at least for finite~$x^+_r$.\\

\section{Conclusions and final comments}\label{secConclusions}

In this work we have explored the unconventional regime $N < 24$ of the one-loop extended semiclassical CGHS model, in contrast to the conventional large-$N$ regime. This line of inquiry was initiated in~\cite{Strominger92} and has more recently served as motivation for introducing unphysical fields and hybrid states in the semiclassical RST model \cite{PSS22, PSS23, PSS23b}. In the present paper we have maintained the original requirement sketched in~\cite{Strominger92}, namely that the unphysical fields (i.e., the Faddeev--Popov ghosts) should not contribute to Hawking radiation. The role of these fields is far from negligible: their contribution to the backreaction is significant. If the total central charge is negative, the curvature singularity is removed as a backreaction effect. The power of backreaction induced by quantum states that  carry no radiation at past/future null infinity was demonstrated explicitly in spherically reduced Einstein gravity, where the event horizon is removed and the backreacted geometry is converted into an asymmetric wormhole with an internal singularity \cite{FFNOS06}; if the sign of the central charge is reversed, both the event horizon and the curvature singularity are removed \cite{dRMN25}. In this paper we studied the natural extension: a dynamical collapse with the $\ket{\mathrm{in}}$ vacuum state. For that, we use the construction of a one-loop extension of the CGHS model that allows analytic solutions. In this setting the singularity is removed, but the event horizon persists.

 The removal of the curvature singularity makes the evolution compatible with unitarity, at least for observers located at finite affine distance from the collapsing matter. This is the most remarkable property of the model, and several comments are in order. A physical signal of unitarity, imprinted in the predicted energy fluxes, is the emergence of bursts of negative energy flux, both at $\mathcal{I}_L^+$ and in the interior of the horizon. This is a characteristic feature of the subcritical regime of the semiclassical CGHS model, below the threshold for black-hole formation \cite{ST94, DM94, CV94, BPP96}. The same point has been emphasized in recent studies on moving mirrors~\cite{Good-Linder-Wilczek}, where the purity of the quantum state is preserved even though the transient negative-energy bursts carry little net energy. Also noteworthy is the large redshift, measured with respect to the affine distance from the collapsing matter, suffered by the energy flux in the interior region, which plays the role of the purification region in our model. In the language of field modes, the redshifted flux in the interior region can be thought of as linked to the Hawking partners, which should therefore undergo extreme stretching during their evolution inside the horizon. This result is also in line with recent findings on moving mirrors~\cite{Ivan} and in quantum gravity~\cite{Ashtekar}.

A further lesson of the present model concerns energy conservation, which proves to be a remarkably subtle issue. This may account for
the fact that energy conservation is not always explicitly addressed in studies of black-hole evaporation, both within the semiclassical
and quantum-gravity approaches. Indeed, to verify energy conservation one requires a field model that is sufficiently tractable at the analytical level, and there are only a few solvable models in which the full quantum evolution of the black hole can be followed. One notable example is the CGHS model itself, with the total negative central charge regime explored in the present work. The possibility of restoring energy conservation while preserving unitarity has been outlined at the end of the previous section. The key idea rests on a natural modification of the model by coupling both chiral sectors, in analogy with moving-mirror models and the behaviour of black holes in four-dimensional gravity. We leave this demanding task for future
work.

{\bf Acknowledgments.}  This work has been supported by Project No. PID2023-149560NB- C21 funded by MCIU /AEI/10.13039/501100011033 / FEDER, UE. The research activities of CGP have also been carried out in the framework of the INFN Research Project QGSKY. JMG is supported by the Ministerio de Ciencia,
Innovaci\'on y Universidades, Ph.D. fellowship, Grant No. FPU22/02528.  The paper is based upon work from COST Action CaLISTA CA21109 supported by COST (European Cooperation in Science and Technology).


\begin{thebibliography}{99}


\bibitem{Hawking76} S.~W.~Hawking,
``Breakdown of Predictability in Gravitational Collapse,'' Phys. Rev. D \textbf{14}, 2460-2473 (1976). 



\bibitem{CGHS}  C.G. Callan, S. B. Giddings, J.A. Harvey, and A. Strominger.  ``Evanescent black holes,''  Phys. Rev. D, \textbf{45}, R1005 (1992).


\bibitem{PSS22} Y.~Potaux, D.~Sarkar and S.~N.~Solodukhin, ``Quantum states and their back-reacted geometries in 2D dilaton gravity,'' Phys. Rev. D \textbf{105}, no.2, 025015 (2022)
[arXiv:2112.03855 [hep-th]].

\bibitem{PSS23}
Y.~Potaux, D.~Sarkar and S.~N.~Solodukhin,
``Spacetime Structure, Asymptotic Radiation, and Information Recovery for a Quantum Hybrid State,''
Phys. Rev. Lett. \textbf{130}, no.26, 261501 (2023)
[arXiv:2212.13208 [hep-th]].

\bibitem{PSS23b} Y.~Potaux, D.~Sarkar and S.~N.~Solodukhin, ``Hybrid quantum states in 2D dilaton gravity,'' Phys. Rev. D \textbf{108}, no.12, 125012 (2023) [arXiv:2310.18745 [hep-th]].


\bibitem{RST}  J.~G.~Russo, L.~Susskind and L.~Thorlacius, ``The Endpoint of Hawking radiation,'' Phys. Rev. D \textbf{46}, 3444-3449 (1992) [arXiv:hep-th/9206070 [hep-th]].


\bibitem{dRMN25} A.~del R{\'\i}o, F.~J.~Mara{\~n}{\'o}n-Gonz{\'a}lez and J.~Navarro-Salas, ``Singularity resolution in spherically reduced 2D semiclassical gravity with negative central charge,''
Phys. Rev. D \textbf{111}, 045025 (2025).



\bibitem{Strominger92} A.~Strominger,
``Faddeev-Popov ghosts and (1+1)-dimensional black hole evaporation,'' Phys. Rev. D \textbf{46}, 4396-4401 (1992).

\bibitem{Polyakov81} A.~M.~Polyakov, ``Quantum Geometry of Bosonic Strings,'' Phys. Lett. B \textbf{103}, 207-210 (1981).

\bibitem{ATV08} A.~Ashtekar, V.~Taveras and M.~Varadarajan, ``Information is Not Lost in the Evaporation of 2-dimensional Black Holes,'' Phys. Rev. Lett. \textbf{100}, 211302 (2008).

\bibitem{APR11} A.~Ashtekar, F.~Pretorius and F.~M.~Ramazanoglu, ``Evaporation of 2-Dimensional Black Holes,'' Phys. Rev. D \textbf{83}, 044040 (2011)
doi:10.1103/PhysRevD.83.044040.


\bibitem{FN05} A. Fabbri and J. Navarro-Salas, {\it Modeling black hole evaporation}, ICP-World Scientific, London (2005)

\bibitem{GSW87} M.~B.~Green, J.~H.~Schwarz and E.~Witten, {\it Superstring theory. Introduction} Vol. 1. Cambridge University Press (1987).


\bibitem{Tong09} D. Tong, ``Lectures on String Theory'', University of Cambridge Part III Mathematical Tripos (Chapter 5, Path integrals and ghosts); arXiv:0908.0333 [hep-th].


\bibitem{Harvey-Strominger92} J. A. Harvey and A. Strominger, ``Quantum Aspects of Black Holes'', hep-th/9209055. 

\bibitem{Strominger} A. Strominger, ``Les Houches Lectures on Black Holes'', hep-th/9501071.


\bibitem{Liberati} S.~Liberati, ``A Real decoupling ghosts quantization of CGHS model for two-dimensional black holes,'' Phys. Rev. D \textbf{51} (1995), 1710-1715
[arXiv:hep-th/9407002 [hep-th]].


\bibitem{CN96} J.~Cruz and J.~Navarro-Salas,
``Solvable models for radiating black holes and area preserving diffeomorphisms,'' Phys. Lett. B \textbf{375}, 47-53 (1996)


\bibitem{Strominger2020} T.~Hartman, E.~Shaghoulian and A.~Strominger, ``Islands in Asymptotically Flat 2D Gravity,'' JHEP \textbf{07}, 022 (2020)
[arXiv:2004.13857 [hep-th]].


\bibitem{Alexandre:2025}
J.~Alexandre, E.~A.~Kontou, D.~Pardo Santos, S.~Pla and A.~Svesko, ``Mass and entropy of asymptotically flat eternal quantum black holes in 2D,''
JHEP \textbf{06}, 040 (2026) [arXiv:2512.19812 [gr-qc]].

\bibitem{BPP} S.~Bose, L.~Parker and Y.~Peleg,
``Semiinfinite throat as the end state geometry of two-dimensional black hole evaporation,'' Phys. Rev. D \textbf{52}, 3512-3517 (1995).




\bibitem{AdW86} A.~Anderson and B.~S.~DeWitt, ``Does the Topology of Space Fluctuate?,'' Found. Phys. \textbf{16}, 91-105 (1986).


\bibitem{ST94} A.~Strominger and L.~Thorlacius, ``Conformally invariant boundary conditions for dilaton gravity,'' Phys. Rev. D \textbf{50}, 5177-5187 (1994)


\bibitem{DM94} S.~R.~Das and S.~Mukherji, ``Boundary dynamics in dilaton gravity,'' Mod. Phys. Lett. A \textbf{9}, 3105-3118 (1994).


\bibitem{CV94} T.~D.~Chung and H.~L.~Verlinde, ``Dynamical moving mirrors and black holes,'' Nucl. Phys. B \textbf{418}, 305-336 (1994).


\bibitem{BPP96} S.~Bose, L.~Parker and Y.~Peleg, ``Hawking radiation and unitary evolution,'' Phys. Rev. Lett. \textbf{76}, 861-864 (1996).




\bibitem{CFTbook}P.~Di Francesco, P.~Mathieu and D.~Senechal, {\it Conformal Field Theory},  Springer-Verlag, New York (1997) 

\bibitem{Kontou2020}
E.~A.~Kontou and K.~Sanders, ``Energy conditions in general relativity and quantum field theory,'' Class. Quant. Grav. \textbf{37} (2020) no.19, 193001 [arXiv:2003.01815 [gr-qc]].



\bibitem{Wilczek93} F. Wilczek, {\it Quantum purity at a small price: easing the black hole paradox}, arXiv:hep-th/9302096 (1993).


\bibitem{FFNOS06} A. Fabbri,  S. Farese, J. Navarro-Salas, G. Olmo and H. Sanchis-Alepuz, {\it Phys. Rev. D} {\bf 73}, 104023 (2006).



\bibitem{Good-Linder-Wilczek} M.~R.~R.~Good, E.~V.~Linder and F.~Wilczek, ``Moving mirror model for quasithermal radiation fields,'' Phys. Rev. D \textbf{101}, 025012 (2020) [arXiv:1909.01129 [gr-qc]].


\bibitem{Ivan}I.~Agullo, P.~Calizaya Cabrera and B.~E.~Navascu{\'e}s,
``Vacuum-purified Hawking radiation from evaporating black holes: Lessons from moving mirrors,''
Phys. Rev. D \textbf{113}, 085010 (2026), 
[arXiv:2512.18354 [gr-qc]].



\bibitem{Ashtekar}A.~Ashtekar,
``Black hole evaporation in loop quantum gravity,''
Gen. Rel. Grav. \textbf{57} (2025) no.2, 48
doi:10.1007/s10714-025-03380-7
[arXiv:2502.04252 [gr-qc]].



 


   
\end{thebibliography}
\end{document}